\documentclass[pra,aps,superscriptaddress,amsmath,amssymb]{revtex4-1}
\usepackage{graphicx}
\usepackage{multirow}
\usepackage{color}
\usepackage{ulem}
\newcommand{\add}[1]{\textcolor{black}{#1}}
\newcommand{\erase}[1]{}
\usepackage{comment}

\begin{document}

\title{Hamiltonian of a flux qubit-LC oscillator circuit in the deep-strong-coupling regime}

\author{F. Yoshihara}
\email{fumiki@nict.go.jp}
\affiliation{Advanced ICT Research Institute, National Institute of Information and Communications Technology, 4-2-1, Nukuikitamachi, Koganei, Tokyo 184-8795, Japan}
\author{S. Ashhab}
\affiliation{Advanced ICT Research Institute, National Institute of Information and Communications Technology, 4-2-1, Nukuikitamachi, Koganei, Tokyo 184-8795, Japan}
\affiliation{Qatar Environment and Energy Research Institute, Hamad Bin Khalifa University, Qatar Foundation, Doha, Qatar}
\author{T. Fuse}
\affiliation{Advanced ICT Research Institute, National Institute of Information and Communications Technology, 4-2-1, Nukuikitamachi, Koganei, Tokyo 184-8795, Japan}
\author{M. Bamba}
\affiliation{Department of Physics, Kyoto University, Kyoto 606-8502, Japan}
\affiliation{PRESTO, Japan Science and Technology Agency, Kawaguchi 332-0012, Japan}
\author{K. Semba}
\email{Present address: Institute for Photon Science and Technology, The University of Tokyo, Tokyo 113-0033, Japan.}
\affiliation{Advanced ICT Research Institute, National Institute of Information and Communications Technology, 4-2-1, Nukuikitamachi, Koganei, Tokyo 184-8795, Japan}

\date{\today}

\begin{abstract}
We derive the Hamiltonian of a superconducting circuit that comprises a single-Josephson-junction flux qubit inductively coupled to an LC oscillator, and we compare the derived circuit Hamiltonian with the quantum Rabi Hamiltonian, which describes a two-level system coupled to a harmonic oscillator.
We show that there is a simple, intuitive correspondence between the circuit Hamiltonian and the quantum Rabi Hamiltonian.
While there is an overall shift of the entire spectrum,
the energy level structure of the circuit Hamiltonian up to the seventh excited states can still be fitted well by the
quantum Rabi Hamiltonian even in the case where the coupling strength is larger than the
frequencies of the qubit and the oscillator, i.e., when the qubit-oscillator circuit is in the deep-strong-coupling regime.
We also show that although the circuit Hamiltonian can be transformed via a unitary transformation to a Hamiltonian containing a capacitive coupling term, the resulting circuit Hamiltonian cannot be approximated by the variant of the quantum Rabi Hamiltonian that is obtained using an analogous procedure for mapping the circuit variables onto Pauli and harmonic oscillator operators, even for relatively weak coupling.
This difference between the flux and charge gauges follows from the properties of the qubit Hamiltonian eigenstates.

\end{abstract}

%PACS numbers:

\maketitle
%intro, derivation of Hamiltonian
\section*{introduction}
Superconducting circuits are one of the most promising platforms for realizing large-scale quantum information processing.
One of the most important features of superconducting circuits is the freedom they allow in their circuit design.
Since the first demonstration of coherent control of a Cooper pair box~\cite{Yasu99Nat}, various types of superconducting circuits have been demonstrated.

The Hamiltonian of a superconducting circuit can be derived using the standard quantization procedure applied to the charge and flux variables in the circuit~\cite{Vool17IJCTA}.
The Hamiltonian of an LC circuit is well known to be that of a harmonic oscillator.
The Hamiltonians of various kinds of superconducting qubits have also been well studied~\cite{Yasu97PRL, Orlando99PRB,Vion02Sci,Martinis02PRL,Koch07PRA}
and these Hamiltonians can be numerically diagonalized to obtain eigenenergies and eigenstates that accurately reproduce experimental data.
On the other hand, the Hamiltonian of circuits containing two or more components, e.g., qubit-qubit, qubit-oscillator, or oscillator-oscillator systems, are usually treated in such a way that the Hamiltonian of the individual components and the coupling among them are separately obtained~\cite{Pashkin2003Nature, Chiorescu04Nat, Blais04PRA, Niskanen06PRBR}.
This separate treatment of individual circuit components works reasonably well in most circuits.
Even for flux qubit-oscillator circuits in the ultrastrong-coupling regime~\cite{Niemczyk10NatP,Forn10PRL}, where the coupling strength $g$ is around 10\% of the oscillator's frequency $\omega$ and the qubit minimum frequency $\Delta_q$, or the deep-strong-coupling regime~\cite{Yoshihara17NatPhys,Yoshihara17PRA,Yoshihara18PRL}, where $g$ is comparable to or larger than $\Delta_q$ and $\omega$,
the experimental data can be well fitted by the quantum Rabi Hamiltonian~\cite{Rabi37PR, Jaynes63IEEE, Braak11PRL},
where a two-level atom and a harmonic oscillator are coupled by a dipole-dipole interaction.
However, the more rigorous approach based on the standard quantization procedure has not been applied to such circuits except in a few specific studies~\cite{Bourassa09PRA,Peropadre2013PRL,Smith2016PRB},
and the validity of describing a flux qubit-oscillator circuit by the quantum Rabi Hamiltonian has been demonstrated in only a few specific circuits~\cite{Manucharyan2017JPA,Bernardis2018PRA}.

\begin{figure}
\includegraphics{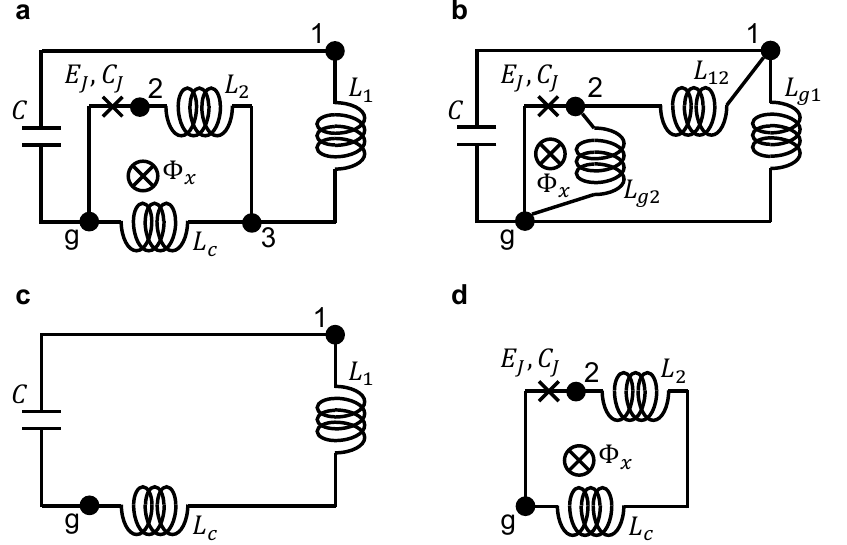}
\caption{Circuit diagrams.
\textbf{a}~A superconducting circuit in which a single-Josephson-junction flux qubit and an LC oscillator are inductively coupled to each other by a shared inductor.
\textbf{b}~Equivalent circuit obtained by applying the so-called Y-$\Delta$ transformation to the inductor network in circuit~\textbf{a}.
\textbf{c}~The outer loop of circuit~\textbf{a}, which forms an LC oscillator.
\textbf{d}~The inner loop of circuit~\textbf{a}, which forms a single-Josephson-junction flux qubit.
}
\label{circ}
\end{figure}

%intro, in this paper
In this paper, we apply the standard quantization procedure to a superconducting circuit in which a single-Josephson-junction flux qubit (an rf-SQUID qubit or a fluxonium-equivalent circuit) and an LC oscillator are inductively coupled to each other by a shared inductor [Fig.~\ref{circ}\textbf{a}], and derive the Hamiltonian of the circuit.
\add{Our model with a single Josephson junction should be sufficient for the purposes of the present study, and the results can be applied to circuits of commonly used multi-Josephson-junction flux qubits~\cite{Chiorescu03Sci,Robertson06PRB}, including the fluxonium.}
Note that single-Josephson-junction flux qubits \add{with different parameters }have been experimentally demonstrated using superinductors, which have been realized by high-kinetic-inductance superconductors~\cite{Peltonen2018SciRep,Hazard2019PRL}, granular aluminum~\cite{Grunhaupt2019NatMat}, and Josephson-junction arrays~\cite{Manucharyan2009Science}. 
The derived circuit Hamiltonian consists of terms associated with the LC oscillator, the flux qubit (and its higher energy levels), and the product of the two flux operators.
Excluding the qubit's energy levels higher than the first excited state,
this circuit Hamiltonian takes the form of the quantum Rabi Hamiltonian, which describes a two-level system coupled to a harmonic oscillator.
%%%
To investigate contributions from the qubit's higher energy levels, we numerically calculate the transition frequencies of the circuit Hamiltonian.
We find that the qubit's higher energy levels mainly cause a negative shift of the entire spectrum, and that the calculated transition frequencies are well fitted by the quantum Rabi Hamiltonian even when the qubit-oscillator circuit is in the deep-strong-coupling regime.
The circuit Hamiltonian can be transformed to one that has a capacitive coupling term by a unitary transformation.
We show, however, that the spectrum of the circuit Hamiltonian cannot be fitted by the variant of the quantum Rabi Hamiltonian that has different Pauli operators in the qubit and coupling terms.
This situation arises when we perform the mapping from circuit variables to Pauli operators for a circuit Hamiltonian that has a capacitive coupling term, i.e. a Hamiltonian expressed in the charge gauge.
We explain the advantage of the flux gauge over the charge gauge in this regard based on the properties of the eigenstates of the flux qubit Hamiltonian.

%figure: circuit diagrams
%end figure

\section*{Results}
\subsection*{circuit Hamiltonian}
\label{Sec_circH}
Following the standard quantization procedure,
nodes are assigned to the circuit as shown in Fig.~\ref{circ}\textbf{a}.
Before deriving the circuit Hamiltonian, the circuit in Fig.~\ref{circ}\textbf{a} is transformed to the one shown in Fig.~\ref{circ}\textbf{b} by applying the so-called Y-$\Delta$ transformation, by which a $\Delta$-shaped network of electrical elements is converted to an equivalent Y-shaped network or vice versa, to the inductor network.
Thus, node 3 surrounded by the inductors is eliminated.
The inductances of the new set of inductors are given as
\begin{eqnarray}
L_{g1} = (L_cL_1+L_cL_2+L_1L_2)/L_2,
\label{Lg1}
\end{eqnarray}
\begin{eqnarray}
L_{g2} = (L_cL_1+L_cL_2+L_1L_2)/L_1,
\label{Lg2}
\end{eqnarray}
and
\begin{eqnarray}
L_{12} = (L_cL_1+L_cL_2+L_1L_2)/L_c.
\label{L12}
\end{eqnarray}

The Lagrangian of the circuit can now be obtained relatively easily~\cite{Vool17IJCTA}:
% Eq: Lagrangian
\begin{eqnarray}
\hat{\mathcal{L}}_{circ}  =  \frac{C}{2}\dot{\hat{\Phi}}_1^2 + \frac{C_J}{2}\dot{\hat{\Phi}}_2^2 +
E_J \cos \left(2\pi\frac{\hat{\Phi}_2-\Phi_x}{\Phi_0}\right)
-\frac{\hat{\Phi}_1^2}{2L_{g1}}-\frac{\hat{\Phi}_2^2}{2L_{g2}}-\frac{(\hat{\Phi}_2-\hat{\Phi}_1)^2}{2L_{12}},
\label{Eq:Lcirc}
\end{eqnarray}
% Eq: end
where $C_J$ and $E_J = I_c\Phi_0/(2\pi)$ are the capacitance and the Josephson energy of the Josephson junction,
$I_c$ is the critical current of the Josephson junction, $\Phi_0 = h/(2e)$ is the superconducting flux quantum,
and $\hat{\Phi}_k$ and $\dot{\hat{\Phi}}_k$ $(k = 1,2)$ are the node flux and its time derivative for node $k$.
The node charges are defined as the conjugate momenta of the node fluxes as $\hat{q}_k = \partial \hat{\mathcal{L}}_{circ}/\partial \dot{\hat{\Phi}}_k$.
After the Legendre transformation, the Hamiltonian is obtained as
% Eq: Hamiltonian
\begin{eqnarray}
\label{Eq:Hcirc}
\hat{\mathcal{H}}_{circ} &=& \frac{\hat{q}_1^2}{2C} + \frac{\hat{q}_2^2}{2C_J} -
E_J \cos \left(2\pi\frac{\hat{\Phi}_2-\Phi_x}{\Phi_0}\right)
+\frac{\hat{\Phi}_1^2}{2L_{LC}}+\frac{\hat{\Phi}_2^2}{2L_{FQ}}-\frac{\hat{\Phi}_1\hat{\Phi}_2}{L_{12}}\\
\nonumber
& = & \hat{\mathcal{H}}_1 + \hat{\mathcal{H}}_2 + \hat{\mathcal{H}}_{12},
\end{eqnarray}
% Eq: end
where
%Eq: H1
\begin{eqnarray}
\hat{\mathcal{H}}_1 & = & \frac{\hat{q}_1^2}{2C} + \frac{\hat{\Phi}_1^2}{2L_{LC}},
\label{Eq:H1}
\end{eqnarray}
% Eq: end
%
%Eq: H2
\begin{eqnarray}
\hat{\mathcal{H}}_2 & = & \frac{\hat{q}_2^2}{2C_J} - E_J \cos \left(2\pi\frac{\hat{\Phi}_2-\Phi_x}{\Phi_0}\right)
+ \frac{\hat{\Phi}_2^2}{2L_{FQ}},
\label{Eq:H2}
\end{eqnarray}
% Eq: end
%
%Eq: H12
\begin{eqnarray}
\hat{\mathcal{H}}_{12} & = & -\frac{\hat{\Phi}_1\hat{\Phi}_2}{L_{12}},
\label{Eq:H12}
\end{eqnarray}
% Eq: end
\begin{eqnarray}
\frac{1}{L_{LC}} & = & \frac{1}{L_{g1}} + \frac{1}{L_{12}},
\label{LLC}
\end{eqnarray}
and
\begin{eqnarray}
\frac{1}{L_{FQ}} & = & \frac{1}{L_{g2}} + \frac{1}{L_{12}}.
\label{LFQ}
\end{eqnarray}

As can be seen from Eq.~(\ref{Eq:Hcirc}), the Hamiltonian $\hat{\mathcal{H}}_{circ}$ can be separated into three parts:
the first part $\hat{\mathcal{H}}_1$ consisting of the charge and flux operators of node 1,
the second part $\hat{\mathcal{H}}_2$ consisting of node 2 operators,
and the third part $\hat{\mathcal{H}}_{12}$ containing the product of the two flux operators.

%%%
\subsection*{separate treatment of the qubit-oscillator circuit}
\label{separate}

\begin{figure}
\includegraphics{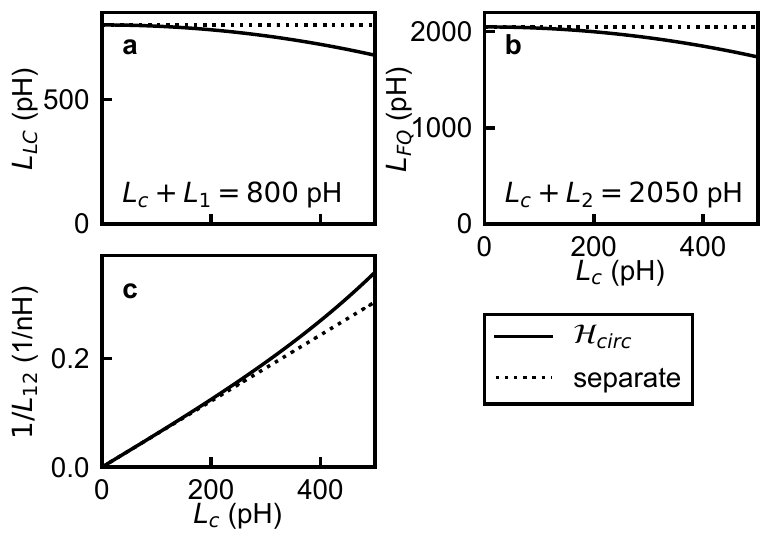}
\caption{%The inductances of the LC oscillator and the flux qubit.
\textbf{a}~The inductances of the LC oscillator $L_{LC}$, \textbf{b}~the inductance of the flux qubit $L_{FQ}$, and \textbf{c}~the inverse inductance of $\hat{\mathcal{H}}_{12}$, $1/L_{12}$ obtained by Eqs.~(\ref{LLC}), (\ref{LFQ}), and (\ref{L12}) (solid lines) are plotted as functions of $L_c$ together with their counterparts in the separate treatment (dotted lines) on condition that the inductance sums are kept constant at $L_c + L_1=800$~pH and $L_c + L_2=2050$~pH.
}
\label{Ls}
\end{figure}

Let us consider \erase{a separate}\add{an alternative} treatment of the circuit shown in Fig.~\ref{circ}\textbf{a}, where the circuit is assumed to be naively divided into two well-defined components.
The capacitor and the inductors in the outer loop of the circuit in Fig.~\ref{circ}\textbf{a}
form an LC oscillator [Fig.~\ref{circ}\textbf{c}].
The Josephson junction and the inductors in the inner loop form a single-Josephson-junction flux qubit [Fig.~\ref{circ}\textbf{d}].
The LC oscillator and the single-Josephson-junction flux qubit share the inductor $L_c$ at the common part of the two loops.
It is now instructive to investigate the relation of the following pairs of Hamiltonians:
$\hat{\mathcal{H}}_1$ and the Hamiltonian of the LC oscillator shown in Fig.~\ref{circ}\textbf{c},
$\hat{\mathcal{H}}_2$ and the Hamiltonian of the flux qubit shown in Fig.~\ref{circ}\textbf{d},
and $\hat{\mathcal{H}}_{12}$ and the Hamiltonian of the inductive coupling between the LC oscillator and the flux qubit
$M\hat{I}_{LC}\hat{I}_{q}=-L_c\hat{\Phi}_1\hat{\Phi}_2/[(L_c + L_1)(L_c + L_2)]$,
where we have used the relations of the oscillator current $\hat{I}_{LC} = \hat{\Phi}_1/(L_c + L_1)$, the qubit current $\hat{I}_q = \hat{\Phi}_2/(L_c + L_2)$, and the mutual inductance $M=L_c$.
Actually, only the inductances are different in the Hamiltonians of the two pictures:
The inductances of the LC oscillator, the flux qubit, and the coupling Hamiltonians derived from the separate treatment are respectively $L_c + L_1$, $L_c + L_2$, and $-(L_c + L_1)(L_c + L_2)/L_c$, while those in $\hat{\mathcal{H}}_{circ}$ are $L_{LC}$, $L_{FQ}$, and $-L_{12}$.
Figure~\ref{Ls} shows the inductances in the separate treatment and in $\hat{\mathcal{H}}_{circ}$ as functions of $L_c$
on condition that the inductance sums are kept constant at $L_c + L_1=800$~pH and $L_c + L_2=2050$~pH.
As $L_c$ approaches 0, or more specifically when $L_c \ll L_1, L_2$, we obtain $L_c + L_1$ $\sim$ $L_{LC}$, $L_c + L_2$ $\sim$ $L_{FQ}$, and $(L_c + L_1)(L_c + L_2)/L_c$ $\sim$ $L_{12}$.
In this way, the Hamiltonian derived from the separate treatment has the same form as $\hat{\mathcal{H}}_{circ}$, and the inductances in the separate treatment approach those of $\hat{\mathcal{H}}_{circ}$.
%end figure

\subsection*{comparison to the quantum Rabi Hamiltonian}
\label{Sec_compHR}
We compare the Hamiltonian $\hat{\mathcal{H}}_{circ}$ with the generalized quantum Rabi Hamiltonian:
%Eq: HR
\begin{eqnarray}
\hat{\mathcal{H}}_{R}/\hbar & = & \omega \left ( \hat{a}^\dagger \hat{a} +\frac{1}{2} \right) -\frac{1}{2}(\varepsilon \hat{\sigma}_x + \Delta_q \hat{\sigma}_z)
+ g\hat{\sigma}_x (\hat{a} + \hat{a}^\dagger)\\
\nonumber
& = & \left ( \hat{\mathcal{H}}_{LC} + \hat{\mathcal{H}}_{FQ} + \hat{\mathcal{H}}_{coup} \right )/\hbar.
\label{Eq:HR}
\end{eqnarray}
% Eq: end
The first part $\hat{\mathcal{H}}_{LC}$ represents the energy of the LC oscillator,
where
$\hat{a}^\dagger$ and $\hat{a}$ are the creation and annihilation operators, respectively.
The second part $\hat{\mathcal{H}}_{FQ}$ represents the energy of the flux qubit written in the energy eigenbasis at $\varepsilon = 0$.
The operators $\hat{\sigma}_{x,z}$ are the standard Pauli operators.
The parameters $\hbar \Delta_q$ and $\hbar \varepsilon$ are the tunnel splitting and the energy bias between the two states with persistent currents flowing in opposite directions around the qubit loop.
The third part $\hat{\mathcal{H}}_{coup}$ represents the inductive coupling energy.

%%%
The relation between $\hat{\mathcal{H}}_1$ and $\hat{\mathcal{H}}_{LC}$ is straightforward.
The resonance frequency and the operators in $\hat{\mathcal{H}}_{LC}$ can be analytically described by the variables and operators in $\hat{\mathcal{H}}_1$ as $\omega = 1/\sqrt{L_{LC}C}$ and $\hat{a} + \hat{a}^\dagger \to \hat{\Phi}_1/(L_{LC}I_{zpf})$,
where $I_{zpf} = \sqrt{\hbar \omega/(2L_{LC})}$ is the zero-point-fluctuation current.
%figure: H2 and HFQ
\begin{figure}
\includegraphics{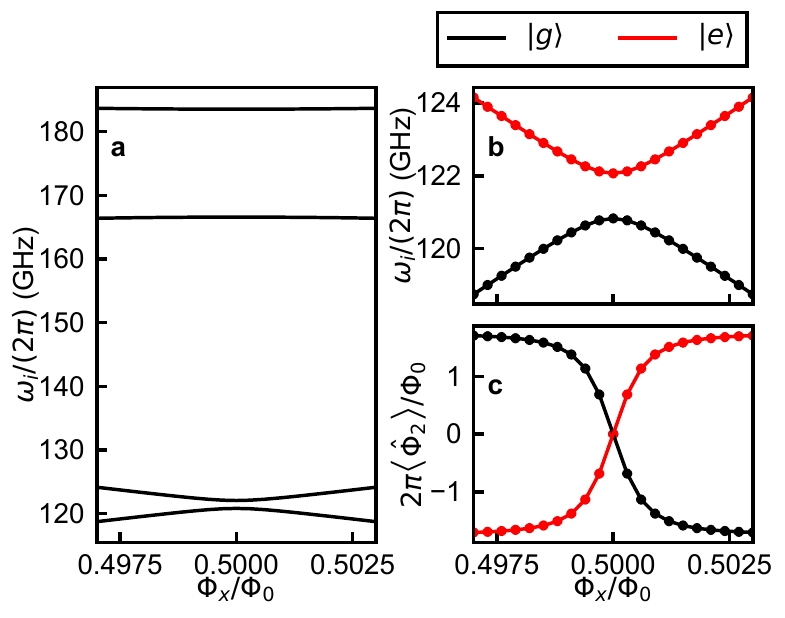}
\caption{\textbf{a}~Numerically calculated energy levels of $\hat{\mathcal{H}}_{2}$ as functions of $\Phi_x$.
\textbf{b}~The lowest two energy levels of $\hat{\mathcal{H}}_{2}$.
\textbf{c}~The expectation values of the flux operator $\left \langle g \left | \hat{\Phi}_2 \right | g \right \rangle$ and $\left \langle e \left | \hat{\Phi}_2 \right | e \right \rangle$.
In \textbf{b}~and \textbf{c}, the solid circles are obtained from numerical calculations of $\hat{\mathcal{H}}_{2}$, while the lines are obtained from fitting by $\hat{\mathcal{H}}_{FQ}$.
The black and red colors respectively indicate states $\left | g\right \rangle$ and $\left | e\right \rangle$.
In the calculation, we used the following parameters: $L_c + L_2=2050$~pH,
$L_J = 990$~pH ($E_J/h=165.1$~GHz), and $C_J=4.84$~fF ($E_C/h = 4.0$~GHz).
}
\label{H2HFQ}
\end{figure}
%end figure
To see the relation between $\hat{\mathcal{H}}_2$ and $\hat{\mathcal{H}}_{FQ}$,
we numerically calculated the eigenenergies of $\hat{\mathcal{H}}_{2}$ as functions of $\Phi_x$.
In the calculation, we used the following parameters: $L_c + L_2=2050$~pH, $L_J = 990$~pH ($E_J/h=165.1$~GHz), and $C_J=4.84$~fF ($E_C/h = 4.0$~GHz).
As shown in Fig.~\ref{H2HFQ}\textbf{a} the lowest two energy levels are well separated from the higher levels, which are more than 40~GHz higher in frequency.
The lowest two energy levels of $\hat{\mathcal{H}}_{2}$ are well approximated by $\hat{\mathcal{H}}_{FQ}$ [Fig.~\ref{H2HFQ}\textbf{b}], which gives almost identical results obtained by the local basis reduction method~\cite{Consani20NJP}, with $\hat{\sigma}_x$ and $\hat{\sigma}_z$ swapped.
The eigen frequencies of the ground state $\omega_0$ and the first excited state $\omega_1$ are respectively fitted by $\omega_{os} - \sqrt{\varepsilon^2 + \Delta_q^2}/2$ and $\omega_{os} + \sqrt{\varepsilon^2 + \Delta_q^2}/2$.
Besides the offset $\omega_{os}$, the fitting parameters are $\Delta_q$ and the maximum persistent current $I_p$,
which is determined as the proportionality constant between the energy bias and the flux bias, $\varepsilon = 2I_p(\Phi_x-0.5\Phi_0)$.
We also numerically calculated the expectation values of the flux operator,
$\left \langle g \left | \hat{\Phi}_2 \right | g \right \rangle$ and $\left \langle e \left | \hat{\Phi}_2 \right | e \right \rangle$, which are well approximated by $\left \langle g \left | \hat{\sigma}_x \right | g \right \rangle$ and $\left \langle e \left | \hat{\sigma}_x \right | e \right \rangle$, respectively [Fig.~\ref{H2HFQ}\textbf{c}].
The expectation values of the flux operator $\left \langle g \left | \hat{\Phi}_2 \right | g \right \rangle$ and $\left \langle e \left | \hat{\Phi}_2 \right | e \right \rangle$ are respectively fitted by $-\Phi_{2max}\varepsilon/\sqrt{\varepsilon^2 + \Delta_q^2}$ and $\Phi_{2max}\varepsilon/\sqrt{\varepsilon^2 + \Delta_q^2}$.
Here, $\left | g\right \rangle$ and $\left | e\right \rangle$ are respectively the ground and excited states of the qubit Hamiltonian $\hat{\mathcal{H}}_{2}$.
The only fitting parameter is determined by the ratio
 $\Phi_{2max} = -\left \langle i \left | \hat{\Phi}_2 \right | i \right \rangle/\left \langle i \left | \hat{\sigma}_x \right | i \right \rangle$ ($i = g,e$).
The Pauli operator $\hat{\sigma}_x$ is therefore identified as being proportional to the flux operator $\hat{\sigma}_x\to-\hat{\Phi}_2/\Phi_{2max}$.
The relation between $\hat{\mathcal{H}}_{12}$ and $\hat{\mathcal{H}}_{coup}$ can now be obtained by using the relations for the oscillator and qubit operators identified above.
% of operators.
This way we find that the Hamiltonian $\hat{\mathcal{H}}_{12}$ can be expressed as
$-(L_{LC}/L_{12})I_{zpf}\Phi_{2max}\hat{\sigma}_x(\hat{a}+\hat{a}^\dagger)$,
which is exactly the same form as $\hat{\mathcal{H}}_{coup}$,
with the coupling strength $\hbar g = -(L_{LC}/L_{12})I_{zpf}\Phi_{2max}$.
Note that $\mathcal{H}_{coup}$ directly derived from the Lagrangian $\hat{\mathcal{L}}_{circ}$ is in the flux gauge, as the qubit-oscillator coupling term is of the form $\hat{\Phi}_1\hat{\Phi}_2$, which is optimal for our system with a single oscillator mode~\cite{Stokes2019NatCom,Roth2019PRR,DiStefano2019NatPhys}.
Excluding the qubit's energy levels higher than the first excited states,
$\hat{\mathcal{H}}_{circ}$ takes the form of $\hat{\mathcal{H}}_{R}$.
In other words, once the circuit parameters, i.e. $\Phi_x$, $L_{c}$, $L_1$, $L_2$, $C$, $E_J$, and $C_J$, are given, the corresponding parameters in $\hat{\mathcal{H}}_{R}$ ($\omega$, $\varepsilon$, $\Delta_q$, and $g$) are set.
The relation between $\hat{\mathcal{H}}_{circ}$ and $\hat{\mathcal{H}}_R$ is summarized in Table~\ref{Rcirc}.

\begin{table}
\begin{tabular}{c c}
    \hline
    \hline
\vspace{0.1cm}
$\hat{\mathcal{H}}_{R}$ & $\hat{\mathcal{H}}_{circ}$\\
    \hline
$\hat{\mathcal{H}}_{LC}$ & $\hat{\mathcal{H}}_{1}$\\
$\hat{\mathcal{H}}_{FQ}$ & $\hat{\mathcal{H}}_{2}$\\
$\hat{\mathcal{H}}_{coup}$ & $\hat{\mathcal{H}}_{12}$\\
$\hat{a}+\hat{a}^\dagger$&$\hat{\Phi}_1/(L_{LC}I_{zpf})$\\
$(\hat{a}-\hat{a}^\dagger)/i$&$\hat{q}_1/(CV_{zpf})$\\
$\hat{\sigma}_x$&$-\hat{\Phi}_2/\Phi_{2max}$\\
$\hat{\sigma}_z$&---\\
$\omega$&$1/\sqrt{L_{LC}C}$\\
$\varepsilon$&$2I_p(\Phi_x-0.5\Phi_0)$\\
$\Delta_q$ & minimum qubit frequency\\
& (numerically evaluated)\\
$g$&$(L_{LC}/L_{12})I_{zpf}\Phi_{2max}/\hbar$\\
    \hline
    \hline
\end{tabular}
\caption{Hamiltonians, operators, and variables in $\hat{\mathcal{H}}_R$ and their counterparts in $\hat{\mathcal{H}}_{circ}$.
$I_{zpf}=\sqrt{\hbar \omega/(2L_{LC})}$ and $V_{zpf}=\sqrt{\hbar \omega/(2C)}$ are the zero-point-fluctuation current and voltage, respectively.
The parameters $\Phi_{2max}$, $I_{p}$, and $\Delta_q$ are obtained by numerically calculating the eigenenergies of $\hat{\mathcal{H}}_2$ as functions of $\Phi_x$ and fitting the lowest two energy levels by $\hat{\mathcal{H}}_{FQ}$.
Note that there is no analytic expression for $\hat{\sigma}_z$ and $\Delta_q$.
}
\label{Rcirc}
\end{table}

\begin{figure}
\includegraphics{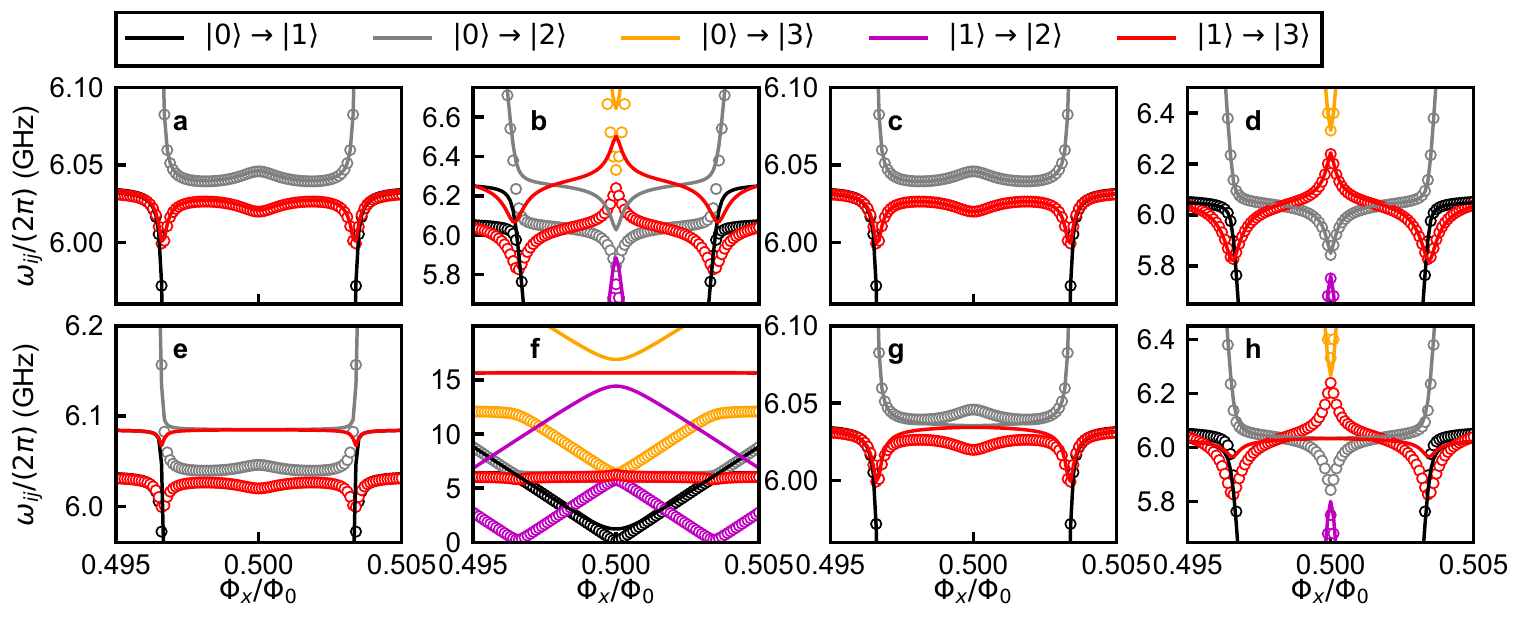}
\caption{
Transition frequencies of the qubit-oscillator circuit fitted with different approximate Hamiltonians.
Numerically calculated transition frequencies from $\hat{\mathcal{H}}_{circ}$ are plotted as circles while the fitting spectra are plotted as lines.
Two mutual inductance values, corresponding to different qubit-oscillator coupling strengths, are used: $L_c$ = 20~pH (\textbf{a}, \textbf{c}, \textbf{e}, \textbf{g}) and $L_c$ = 350~pH (\textbf{b}, \textbf{d}, \textbf{f}, \textbf{h}).
The spectra are fitted with $\hat{\mathcal{H}}_{R}$ (\textbf{a}, \textbf{b}, \textbf{c}, \textbf{d}) and $\hat{\mathcal{H}}_{R}'$ (\textbf{e}, \textbf{f}, \textbf{g}, \textbf{h}).
\textbf{a}, \textbf{b}~The parameters of $\hat{\mathcal{H}}_{R}$ are obtained from the relations described in Table~\ref{Rcirc}: $\omega/(2\pi) = 6.033$~GHz, $\Delta_q/(2\pi) = 1.240$~GHz, $g/(2\pi)=0.424$~GHz, and $I_p = 281.3$~nA for $L_c = 20$~pH and
$\omega/(2\pi) = 6.272$~GHz, $\Delta_q/(2\pi) = 2.139$~GHz, $g/(2\pi)=7.338$~GHz, and $I_p = 282.5$~nA for $L_c = 350$~pH.
\textbf{c}, \textbf{d}~The parameters of $\hat{\mathcal{H}}_{R}$ are obtained by fitting the spectra of $\hat{\mathcal{H}}_{circ}$ to $\hat{\mathcal{H}}_{R}$: $\omega/(2\pi) = 6.033$~GHz, $\Delta_q/(2\pi) = 1.240$~GHz, $g/(2\pi)=0.430$~GHz, and $I_p = 281.3$~nA for $L_c = 20$~pH and
$\omega/(2\pi) = 6.064$~GHz, $\Delta_q/(2\pi) = 2.388$~GHz, $g/(2\pi)=7.822$~GHz, and $I_p = 282.9$~nA for $L_c = 350$~pH.
\textbf{e}, \textbf{f}~The parameters of $\hat{\mathcal{H}}_{R}'$ are obtained from the circuit parameters of $\hat{\mathcal{H}}_{circ}'$ in the similar way as in Table~\ref{Rcirc}: $\omega/(2\pi) = 6.085$~GHz, $\Delta_q/(2\pi) = 1.238$~GHz, $g'/(2\pi)=$0.043~GHz, and $I_p = 281.3$~nA for $L_c = 20$~pH,
and $\omega/(2\pi) = 15.66$~GHz, $\Delta_q/(2\pi) = 1.238$~GHz, $g'/(2\pi)=$0.492~GHz, and $I_p = 281.3$~nA for $L_c = 350$~pH.
\textbf{g}, \textbf{h}~The parameters of $\hat{\mathcal{H}}_{R}'$ are obtained by fitting the spectra of $\hat{\mathcal{H}}_{circ}'$ to $\hat{\mathcal{H}}_{R}'$:
$\omega/(2\pi) = 6.035$~GHz, $\Delta_q/(2\pi) = 1.220$~GHz, $g/(2\pi)=0.089$~GHz, and $I_p = 281.3$~nA for $L_c = 20$~pH and
$\omega/(2\pi) = 6.034$~GHz, $\Delta_q/(2\pi) = 0.233$~GHz, $g/(2\pi)=0.165$~GHz, and $I_p = 281.4$~nA for $L_c = 350$~pH.
Black, gray, orange, magenta, and red colors indicate transitions $\left | 0\right \rangle \to \left | 1\right \rangle$, $\left | 0\right \rangle \to \left | 2\right \rangle$, $\left | 0\right \rangle \to \left | 3\right \rangle$, $\left | 1\right \rangle \to \left | 2\right \rangle$, and $\left | 1\right \rangle \to \left | 3\right \rangle$, respectively.
}
\label{wij_phix}
\end{figure}

\begin{figure}
\includegraphics{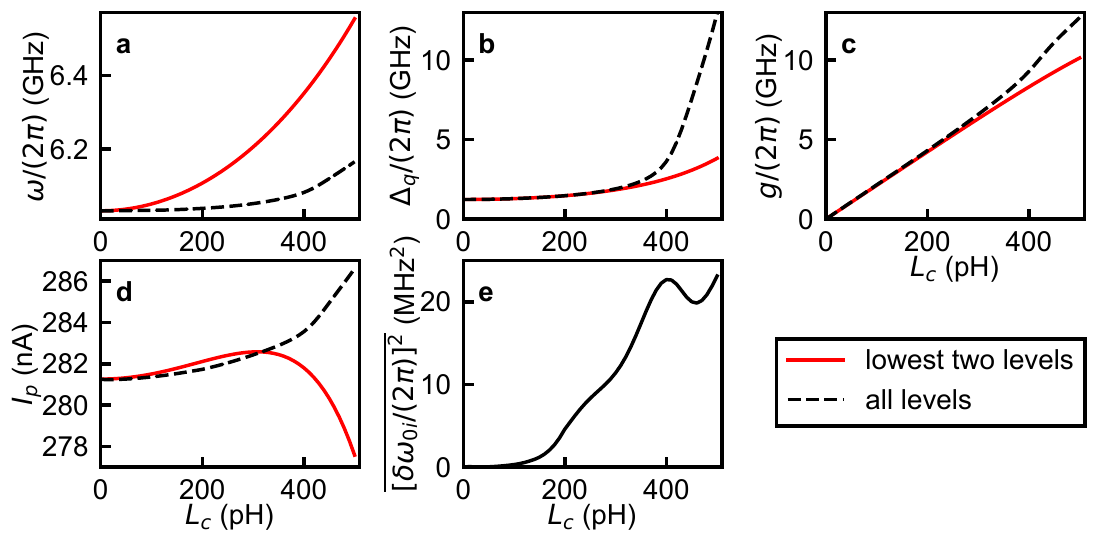}
\caption{
Parameters of $\hat{\mathcal{H}}_{R}$ obtained from the relations described in Table~\ref{Rcirc}, (red solid lines) and by fitting the spectra of $\hat{\mathcal{H}}_{circ}$ up to the third excited state
to $\hat{\mathcal{H}}_{R}$ (black dashed lines) as a function of $L_c$.
Panels \textbf{a}-\textbf{d}~respectively correspond to the oscillator frequency $\omega$, qubit frequency $\Delta_q$,
coupling strength $g$, and persistent current $I_p$ in $\hat{\mathcal{H}}_R$.
\textbf{e}~The average of the squares of the residuals in the least-squares method for obtaining the parameters of $\hat{\mathcal{H}}_{R}$ by fitting, $\overline{[\delta \omega_{0i}/(2\pi)]^2}$ ($i$ = 1, 2, and 3).
}
\label{circR_LC}
\end{figure}

%%%
As previously mentioned, $\hat{\mathcal{H}}_{R}$ considers only the lowest two energy levels of the flux qubit,
while $\hat{\mathcal{H}}_{circ}$ includes all energy levels.
To investigate the effect of the qubit's higher energy levels, we perform numerical calculations and compare the energy levels calculated by $\hat{\mathcal{H}}_{R}$ and $\hat{\mathcal{H}}_{circ}$.
The details of the numerical diagonalization of $\hat{\mathcal{H}}_{circ}$ are given in \textbf{methods}. %Appendix~\ref{NC}.
Since the contributions from the qubit's higher energy levels are expected to become larger as the coupling strength increases,
we consider parameters that cover a wide range of coupling strengths from the weak-coupling to the deep-strong-coupling regime.
In the calculations, we fix $L_c+L_1=800$~pH, $L_c + L_2=2050$~pH, $C=0.87$~pF, $L_J = 990$~pH ($E_J/h=165.1$~GHz), and $C_J=4.84$~fF ($E_C/h = 4.0$~GHz), and sweep the flux bias $\Phi_x$ around $\Phi_0/2$ at various values of $L_c$.
These parameters are used in all the calculations for Figs.~\ref{wij_phix}-\ref{adaga}.
Some of our calculations were performed using the QuTiP simulation package~\cite{Johansson13CPC}.

%figure: wij_nf
%end figure
Transition frequencies of the qubit-oscillator circuit $\omega_{ij}$ corresponding to the transition $\left | i \right \rangle \to \left | j \right \rangle$ numerically calculated from $\hat{\mathcal{H}}_{circ}$ around the resonance frequency of the oscillator $\omega$ are plotted in Fig.~\ref{wij_phix} for \textbf{a} $L_c=20$~pH and \textbf{b} $L_c=350$~pH, where the indices $i$ and $j$ label the energy eigenstates according to their order in the energy-level ladder, with the index 0 denoting the ground state.
The same circuit parameters as in Fig.~\ref{H2HFQ} are used.
In the case $L_c=20$~pH, two characteristic features are observed: avoided level crossings between the qubit and oscillator transition signals approximately at $\Phi_x/\Phi_0 = 0.497$ and 0.503, and the dispersive shift that for example creates the separation between the frequencies of the transitions $\left | 0\right \rangle \to \left | 2\right \rangle$ and $\left | 1\right \rangle \to \left | 3\right \rangle$, leading to the peak/dip feature at the symmetry point, i.e. $\Phi_x/\Phi_0 = 0.5$.
Note that transitions $\left |0\right \rangle \to \left |2\right \rangle$ and $\left |1\right \rangle \to \left |3\right \rangle$ around $\Phi_x/\Phi_0 = 0.5$ respectively correspond to transitions $\left |g0\right \rangle \to \left |g1\right \rangle$ and $\left |e0\right \rangle \to \left |e1\right \rangle$, where ``$g$" and ``$e$" denote, respectively,
the ground and excited states of the qubit, and ``0" and ``1" the number of photons in the oscillator's Fock state.
In the case $L_c=350$~pH, the characteristic spectrum indicates that the qubit-oscillator circuit is in the deep-strong-coupling regime~\cite{Yoshihara17PRA}.
Transition frequencies of $\hat{\mathcal{H}}_{R}$ are also plotted in Figs.~\ref{wij_phix}\textbf{a} and \textbf{b}.
It should be mentioned that the parameters of $\hat{\mathcal{H}}_{R}$ are obtained in two different ways. Here, the parameters of $\hat{\mathcal{H}}_{R}$ are obtained from the relations described in Table~\ref{Rcirc}.
The overall shapes of the spectra of $\hat{\mathcal{H}}_{R}$ and $\hat{\mathcal{H}}_{circ}$ look similar.
On the other hand, the shift of the entire spectrum becomes as large as more than 200~MHz for $L_c = 350$~pH.
To quantify the difference of the spectra between $\hat{\mathcal{H}}_{R}$ and $\hat{\mathcal{H}}_{circ}$,
transition frequencies up to the third excited state numerically calculated from $\hat{\mathcal{H}}_{circ}$ are fitted by $\hat{\mathcal{H}}_{R}$.
In the fitting, $\hat{\mathcal{H}}_{R}$ with an initial approximate set of parameters is numerically diagonalized and then the parameters, $\omega$, $\Delta_q$, $g$, and $I_p (=\varepsilon/[2(\Phi_x-0.5\Phi_0)])$, are varied to obtain the best fit.
Figures~\ref{wij_phix}\textbf{c} and \textbf{d} show that the same spectra of $\hat{\mathcal{H}}_{circ}$ in Figs.~\ref{wij_phix}\textbf{a} and \textbf{b} are well fitted by $\hat{\mathcal{H}}_{R}$, but with a different parameter set.

%figure: params circ R_LC
%end figure

Figure~\ref{circR_LC} shows parameters of $\hat{\mathcal{H}}_{R}$ obtained from the relations described in Table~\ref{Rcirc}, and by fitting the spectra of $\hat{\mathcal{H}}_{circ}$ up to the third excited state
to $\hat{\mathcal{H}}_{R}$ as a function of $L_c$.
The parameters of $\hat{\mathcal{H}}_{R}$ obtained in these two different ways become quite different from each other at large values of $L_c$ due to contributions of the qubit's higher energy levels.
Interestingly, the average of the squares of the residuals in the least-squares method for obtaining the parameters of $\hat{\mathcal{H}}_{R}$ by fitting, $\overline{[\delta \omega_{0i}/(2\pi)]^2}$ ($i$ = 1, 2, and 3),
remains rather small, at most $25~$MHz$^2$,
which is consistent with the good fitting shown in Figs.~\ref{wij_phix}\textbf{c} and \textbf{d}.
In fact, the energy level structure of $\hat{\mathcal{H}}_{circ}$ up to the seventh excited state can still be fitted well by the quantum Rabi Hamiltonian as shown in Supplementary Information~\cite{SI}.

\subsection*{Hamiltonian in the charge gauge}
\label{gauge_trans}

The circuit Hamiltonian $\hat{\mathcal{H}}_{circ}$ is in the flux gauge,
as the qubit-oscillator coupling term is of the form $\hat{\Phi}_1\hat{\Phi}_2$.
The Hamiltonian can be transformed into the charge gauge:
\begin{eqnarray}
\nonumber
\hat{\mathcal{H}}_{circ}' &=& \hat{\mathcal{U}}^\dagger \hat{\mathcal{H}}_{circ} \hat{\mathcal{U}}\\
\label{Hcircp}
&=&\left (\frac{1}{2C} + \frac{L_{LC}^2}{2C_JL_{12}^2}\right )\hat{q}_1^2 + \frac{\hat{\Phi}_1^2}{2L_{LC}} + \frac{\hat{q}_2^2}{2C_J}
 + \frac{1}{2}\left ( \frac{1}{L_{FQ}}-\frac{L_{LC}}{L_{12}^2}\right ) \hat{\Phi}_2^2
 - E_J\cos \left ( 2\pi \frac{\hat{\Phi}_2-\Phi_x}{\Phi_0} \right ) - \frac{L_{LC}}{C_JL_{12}}\hat{q}_1\hat{q}_2\\
\nonumber
&=& \hat{\mathcal{H}}_{1}' + \hat{\mathcal{H}}_{2}' + \hat{\mathcal{H}}_{12}',
\end{eqnarray}
where
\begin{eqnarray}
\hat{\mathcal{U}} &=& \exp \left ( \frac{1}{i\hbar}\frac{L_{LC}}{L_{12}}\hat{\Phi}_2\hat{q}_1\right ),\\
\hat{\mathcal{H}}_{1}' &=&\left (\frac{1}{2C} + \frac{L_{LC}^2}{2C_JL_{12}^2}\right )\hat{q}_1^2 + \frac{\hat{\Phi}_1^2}{2L_{LC}},\\
%\label{H1p}
\hat{\mathcal{H}}_{2}' &=&\frac{1}{2C_J}\hat{q}_2^2 + \frac{1}{2}\left ( \frac{1}{L_{FQ}}-\frac{L_{LC}}{L_{12}^2}\right ) \hat{\Phi}_2^2
- E_J\cos \left ( 2\pi \frac{\hat{\Phi}_2-\Phi_x}{\Phi_0} \right ),\\
%\label{H2p}
\hat{\mathcal{H}}_{12}' &=&- \frac{L_{LC}}{C_JL_{12}}\hat{q}_1\hat{q}_2.
%\label{H12p}
\end{eqnarray}
The details of the calculations are given in Supplementary Information~\cite{SI}.
As can be seen in Eq.~(\ref{Hcircp}), the Hamiltonian $\hat{\mathcal{H}}_{circ}'$ can be separated into three parts:
the first part $\hat{\mathcal{H}}_1'$ consisting of the charge and flux operators of node 1,
the second part $\hat{\mathcal{H}}_2'$ consisting of the charge and flux operators of node 2,
and the third part $\hat{\mathcal{H}}_{12}'$ containing the product of the two charge operators.
It is worth mentioning that the inductance in $\hat{\mathcal{H}}_{2}'$ is equal to $L_c + L_2$,
which is the inductance of the flux qubit in the separate treatment:

\begin{eqnarray}
\nonumber
\frac{1}{L_{FQ}}-\frac{L_{LC}}{L_{12}^2} & = & \frac{1}{L_{12}} + \frac{1}{L_{g2}}-\frac{L_{g1}}{L_{12}(L_{12}+L_{g1})}\\
\nonumber
&=& \frac{L_c+L_1}{L_cL_1+L_cL_2+L_1L_2}
- \frac{L_c^2}{(L_cL_1+L_cL_2+L_1L_2)(L_c+L_2)}\\
\nonumber
&=& \frac{(L_c+L_1)(L_c+L_2)-L_c^2}{(L_cL_1+L_cL_2+L_1L_2)(L_c+L_2)}\\
& = & \frac{1}{L_c+L_2}.
\label{Eq:HR}
\end{eqnarray}

\begin{figure}
\includegraphics{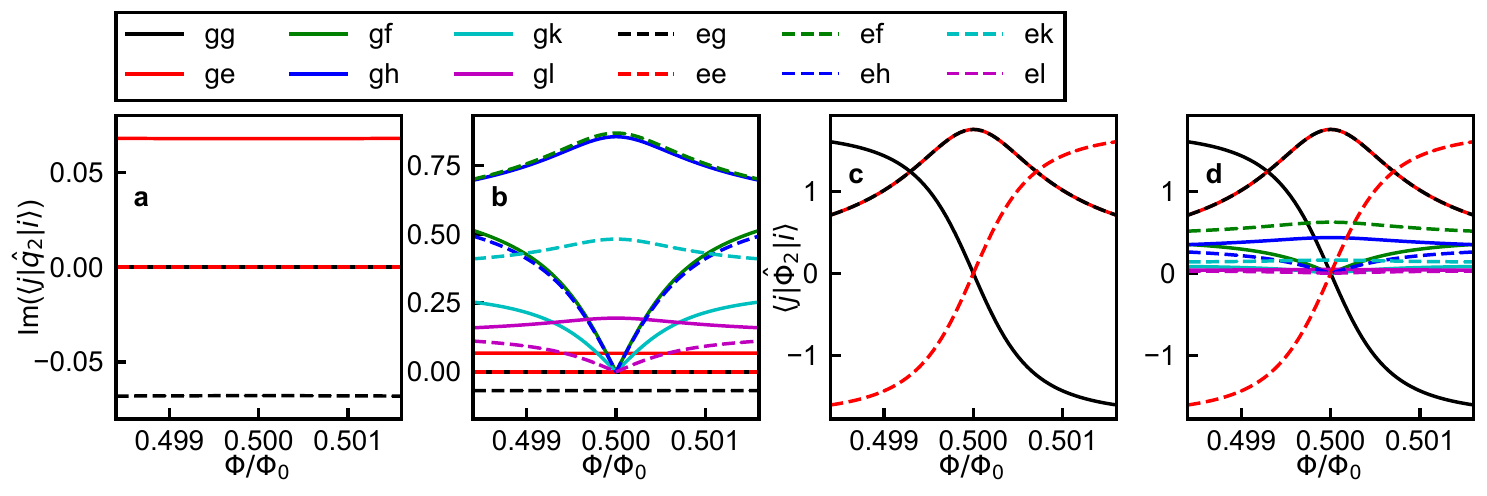}
\caption{
Numerically calculated matrix elements $\left \langle j \left | \hat{q}_2 \right | i \right \rangle$ for $i = g, e$ and \textbf{a}~$j = g, e$, \textbf{b}~$j = g, e, f, h, k, l$, where $\left | f \right \rangle$, $\left | h \right \rangle$, $\left | k \right \rangle$, and $\left | l \right \rangle$ respectively represent the second, third, fourth, and fifth excited states of $\hat{\mathcal{H}}_2$.
%Note that $\hat{\mathcal{H}}_2' = \hat{\mathcal{H}}_2$ in the case $L_c = 0$.
Numerically calculated matrix elements $\left \langle j \left | \hat{\Phi}_2 \right | i \right \rangle$ for $i = g, e$ and \textbf{c}~$j = g, e$, \textbf{d}~$j = g, e, f, h, k, l$.
Note that the x axis ranges are smaller than those in Fig. \ref{wij_phix}.
}
\label{mijc}
\end{figure}
The overall form of $\hat{\mathcal{H}}'_{circ}$ [Eq.~(\ref{Hcircp})] is almost the same as that of $\hat{\mathcal{H}}_{circ}$ [Eq.~(\ref{Eq:Hcirc})], with the exception that the coupling term is of the form $\hat{q}_1\hat{q}_2$ instead of $\hat{\Phi}_1 \hat{\Phi}_2$.
When we examined the mapping between $\hat{\mathcal{H}}_{2}$ and $\hat{\mathcal{H}}_{FQ}$, we explained
that the operator $\hat{\Phi}_2$ can be identified as $\hat{\sigma}_x$.
Similarly, if we calculate the matrix elements of the operator $\hat{q}_2$ for the two lowest qubit states,
we find that $\left \langle g \left | \hat{q}_2 \right | g \right \rangle = \left \langle e \left | \hat{q}_2 \right | e \right \rangle = 0$ and $\left \langle g \left | \hat{q}_2 \right | e \right \rangle = -\left \langle e \left | \hat{q}_2 \right | g \right \rangle \neq 0$ [Fig.~\ref{mijc}\textbf{a}].
We can therefore identify the operator $\hat{q}_2$ as the qubit operator $\hat{\sigma}_y$.

It then seems natural to look for a mapping of $\hat{\mathcal{H}}'_{circ}$ to the variant of the quantum Rabi Hamiltonian:
\begin{eqnarray}
\hat{\mathcal{H}}_{R}'/\hbar & = & \omega \left ( \hat{a}^\dagger \hat{a} +\frac{1}{2} \right) -\frac{1}{2}(\varepsilon \hat{\sigma}_x + \Delta_q \hat{\sigma}_z)
+ ig'\hat{\sigma}_y (\hat{a} - \hat{a}^\dagger)\\
\nonumber
& = & \left ( \hat{\mathcal{H}}_{LC} + \hat{\mathcal{H}}_{FQ} + \hat{\mathcal{H}}_{coup}' \right )/\hbar.
\label{Eq:HR}
\end{eqnarray}
Compared with $\hat{\mathcal{H}}_{R}$, only $\hat{\mathcal{H}}_{coup}$ is replaced by $\hat{\mathcal{H}}_{coup}'$.
It should be noted, however, that there is a physical difference between the two Hamiltonians and that there is no simple mapping between them.
The operator $\hat{\sigma}_x$ in $\hat{\mathcal{H}}_{coup}$ is the same as one of the Pauli operators in $\hat{\mathcal{H}}_{FQ}$.
In contrast, the operator $\hat{\sigma}_y$ in $\hat{\mathcal{H}}_{coup}'$ is different from the two operators ($\hat{\sigma}_x$ and $\hat{\sigma}_z$) in $\hat{\mathcal{H}}_{FQ}$.
As a result, $\hat{\mathcal{H}}_{R}$ and $\hat{\mathcal{H}}'_{R}$ are physically different and for example produce different spectra.

In Fig.~\ref{wij_phix} we plot a few of the transition frequencies in the circuit's spectrum along with the corresponding transition frequencies obtained from $\hat{\mathcal{H}}_{R}'$.
When the parameters of $\hat{\mathcal{H}}_{R}'$ are read off $\hat{\mathcal{H}}_{circ}'$, similarly to what is shown in Table I,
Figs.~\ref{wij_phix}\textbf{e} and \ref{wij_phix}\textbf{f} show that the fitting is poor in most parts of the spectrum, even for the relatively weak coupling case $L_c=20$~pH.
Here, $g' = q_{1zpf}q_{2max}L_{LC}/(C_JL_{12})$, $q_{1zpf} =\sqrt{\hbar \omega'C'/2}$, $\omega' = 1/\sqrt{L_{LC}}\times\sqrt{(1/C)+(L_{LC}^2/C_JL_{12}^2)}$, $1/C' = (1/C) + [L_{LC}^2/(C_JL^2_{12})]$, and $q_{2max}$ is numerically calculated as shown in Fig. \ref{mijc}\textbf{a}.
Contrary to the open grey and red circles obtained from $\hat{\mathcal{H}}_{circ}'$, the solid grey and red lines obtained from $\hat{\mathcal{H}}_{R}'$ are larger and do not show the peaks and dips around the symmetry point.
Even with a numerical optimization of the fitting parameters [Figs.~\ref{wij_phix}\textbf{g} and \ref{wij_phix}\textbf{h}], only parts of the spectrum can be fitted well.
In particular, the peaks and dips that occur in the spectrum at $\Phi_x/\Phi_0=0.5$ are not reproduced by $\hat{\mathcal{H}}_{R}'$.

Here it is useful to consider the two characteristic features in the spectrum, i.e. the avoided crossings and dispersive shifts, as being the result of energy shifts in the spectrum of an uncoupled circuit when the coupling term is added.
The gap of an avoided level crossing, or in other words the Rabi splitting, is proportional to the matrix element of the coupling term between the relevant energy eigenstates of the uncoupled system.
The dispersive shift of one energy level caused by another energy level is proportional to the square of the matrix element between the two energy eigenstates according to perturbation theory.
The details of the dispersive shifts up to second order in perturbation theory are described in Supplementary information~\cite{SI}. %Appendix~\ref{perturb}.

The difference between the spectra shown in the solid lines in Figs.~\ref{wij_phix}\textbf{a} and \ref{wij_phix}\textbf{e} is attributed to the difference between the matrix elements of the qubit's flux and charge operators, $\hat{\Phi}_2$ and $\hat{q}_2$.
The flux bias dependences of the numerically calculated matrix elements of the qubit's charge and flux operators are shown in Fig.~\ref{mijc}.
Regarding the matrix elements of the qubit's flux operator $\left | \left \langle j\left | \hat{\Phi}_2 \right | i\right \rangle \right |$ ($i = g,e$), those involving the higher qubit levels $j = f, h, k, l$ are smaller than those of $j = g, e$.
Here, $\left | f \right \rangle$, $\left | h \right \rangle$, $\left | k \right \rangle$, and $\left | l \right \rangle$ respectively represent the second, third, fourth, and fifth excited states of $\hat{\mathcal{H}}_2$.
Regarding the matrix elements of the qubit's charge operator $\left | \left \langle j\left | \hat{q}_2 \right | i\right \rangle \right |$ ($i = g,e$), on the other hand, some of those involving the higher qubit levels $j = f, h, k, l$ are significantly larger than those of $j = g, e$ in most of the flux bias range.
%Here, $\left | f' \right \rangle$, $\left | h' \right \rangle$, $\left | k' \right \rangle$, and $\left | l' \right \rangle$ respectively represent the second, the third, the fourth, and the fifth excited states of $\hat{\mathcal{H}}_2'$.
This difference stems from the fact that the qubit Hamiltonian involves a double-well potential of the flux variables, as explained in Supplementary information~\cite{SI}.

The matrix elements $\left \langle g \left | \hat{\Phi}_2 \right | e \right \rangle$ and $\left \langle e \left | \hat{\Phi}_2 \right | g \right \rangle$ have a peak at $\Phi/\Phi_0 = 0.5$, which directly leads to the peak/dip feature at the symmetry point as shown in the solid lines in Fig.~\ref{wij_phix}\textbf{a}.
On the other hand, the matrix elements $\left \langle g \left | \hat{q}_2 \right | e \right \rangle$ and $\left \langle e \left | \hat{q}_2 \right | g \right \rangle$ are almost constant in the flux bias range of Fig.~\ref{mijc}, which is consistent with the absence of a peak/dip feature at the symmetry point in the solid lines in Fig.~\ref{wij_phix}\textbf{e}.
Instead, the matrix elements $\left \langle h \left | \hat{q}_2 \right | g \right \rangle$ and $\left \langle f \left | \hat{q}_2 \right | e \right \rangle$, which have somewhat similar flux-bias dependence to those of $\left \langle g \left | \hat{\Phi}_2 \right | e \right \rangle$ and $\left \langle e \left | \hat{\Phi}_2 \right | g \right \rangle$, have peaks at $\Phi/\Phi_0 = 0.5$.
As a result, for the flux gauge, the contribution from higher levels is just a small correction, while it cannot be neglected in the charge gauge.
For this reason, the flux gauge turns out to be more convenient for purposes of mapping $\hat{\mathcal{H}}_{circ}$ to the quantum Rabi Hamiltonian.

It is worth mentioning that we are dealing with an inductively coupled circuit, and one might think that the nature of the coupling, i.e. inductive or capacitive, will determine which gauge is more suitable, especially because the two gauges differ mainly by the form of the coupling term.
We show that the deciding factor is the qubit rather than the coupling.
As a result, if we have a flux qubit capacitively coupled to the oscillator, the flux gauge will be more suitable for the purpose of approximating the circuit Hamiltonian by the quantum Rabi Hamiltonian.
\add{It is also worth mentioning that our results should apply to fluxonium-resonator circuits, since the fluxonium Hamiltonian close to the degeneracy point also involves a double-well potential of the flux variables.
It should be noted, however, that the fluxonium's transition frequencies to the higher energy levels, $\omega_{02}$ and $\omega_{03}$, are smaller than those of flux qubits, and, hence, the contribution from higher levels would be larger in fluxonium-resonator circuits.}

We note here that the differences between the energy spectra obtained from $\hat{\mathcal{H}}_{circ}$, $\hat{\mathcal{H}}_{R}$, and $\hat{\mathcal{H}}_{R}'$ in the case $\varepsilon = 0$, $\Delta_q = \omega$, and a quadratic-plus-quartic potential energy function were extensively studied in Ref.~\cite{Bernardis2018PRA}.
Contrary to the case of $\Delta_q < \omega$, the spectra obtained from $\hat{\mathcal{H}}_{circ}$ and $\hat{\mathcal{H}}_{R}$ are almost identical for the lowest energy levels in the whole range of $g/\omega$ while that of $\hat{\mathcal{H}}_{R}'$ deviates from the other two when $g/\omega \gtrsim 0.1$.
The Rabi splitting between the first two excited energy levels, which is symmetric and is proportional to $g$, is clearly visible for $g/\omega \ll 0.1$, while the dispersive shift due to the higher energy levels is proportional to $g^2$, and can be observed only when $g/\omega \gtrsim 0.1$.
We also note that for $L_c=20$~pH, $g'/(2\pi) = 0.043$~GHz, which is much smaller than $g/(2\pi) = 0.424$~GHz.
On the other hand, the gap of the avoided level crossings plotted as lines in Fig.~\ref{wij_phix}\textbf{e} is not much smaller than that in Fig.~\ref{wij_phix}\textbf{a} considering  $g' \sim 0.1g$.
The gap of the avoided level crossings for $\hat{\mathcal{H}}_{R}'$ is $2g'$ while that for $\hat{\mathcal{H}}_{R}$ is $2g\Delta_q/\omega$.
The factor $\Delta_q/\omega = 0.206$ partly explains why the avoided level crossings for $\hat{\mathcal{H}}_{R}$ and $\hat{\mathcal{H}}_{R}'$ are not very different from each other.

\subsection*{expectation values of the photon number and the field operator}
\label{expectation}
%\begin{figure*}[t]
% \begin{center}
%  \includegraphics{fig.eps}
%  \caption{caption}
%  \label{}
% \end{center}
%\end{figure*}
\begin{figure*}
\begin{center}
\includegraphics{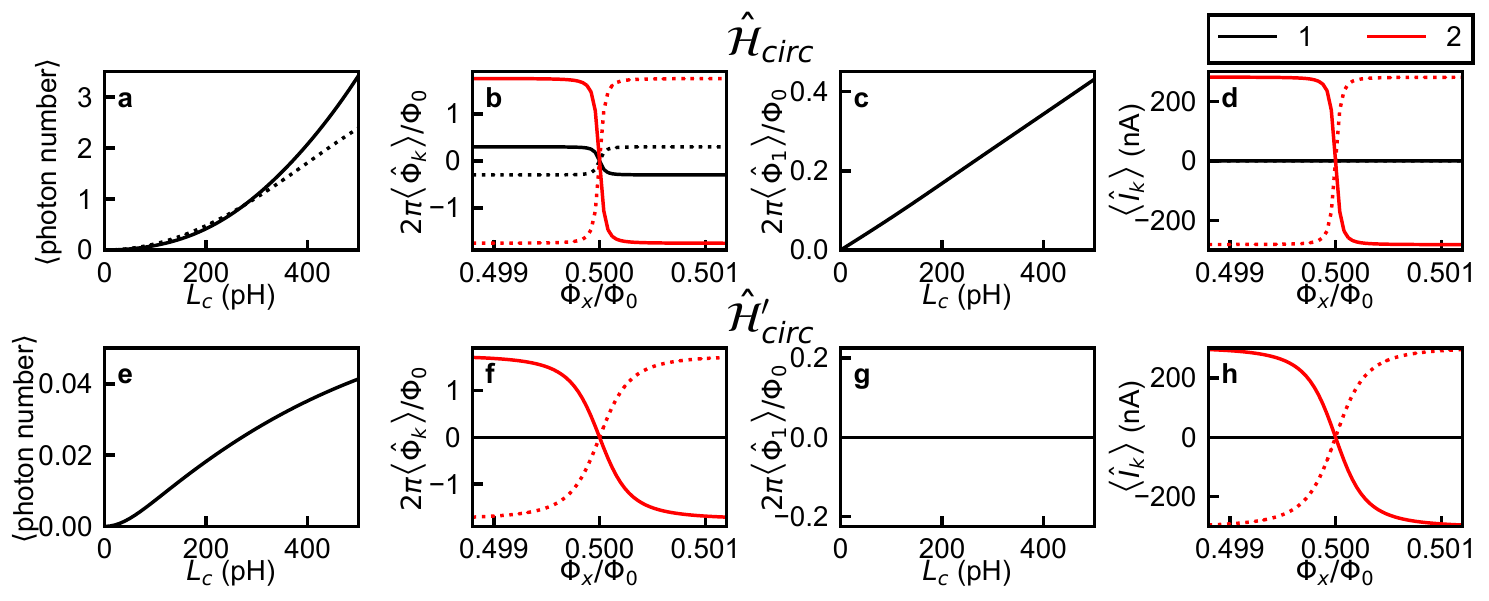}
\caption{
Expectation values of photon number, flux, and current operators for \textbf{a}-\textbf{d}~$\hat{\mathcal{H}}_{circ}$ and for \textbf{e}-\textbf{f}~$\hat{\mathcal{H}}_{circ}'$.
\textbf{a}, \textbf{e}~Expectation numbers of photons in the oscillator in the ground state as a function of $L_c$.
The solid curve corresponds to $\left \langle 0 \left |\hat{\mathcal{H}}_1/\hbar\right | 0 \right \rangle/(1/\sqrt{L_{LC}C})-0.5$, while the dashed line in panel~\textbf{a} indicates the simple expression $(g/\omega)^2$, which is valid in the limit $\Delta_q \ll \omega$.
\textbf{b}, \textbf{f} [\textbf{d}, \textbf{h}]~Expectation values of fluxes $2\pi \left \langle \hat{\Phi}_k\right \rangle/\Phi_0$ [Expectation values of currents $\left \langle \hat{I}_k \right \rangle$] as functions of the flux bias $\Phi_x$ in the case $L_c=350$~pH.
Black and red colors respectively indicate $k = 1$ and 2,
and solid and dashed lines indicate the ground and the excited states, respectively.
\textbf{c}, \textbf{g}~Expectation value of flux $2\pi \left \langle \hat{\Phi}_1\right \rangle/\Phi_0$ in the ground state as a function of $L_c$ at $\Phi_{x}/\Phi_{0} = 0.498$.
}
\label{adaga}
\end{center}
\end{figure*}
%end figure
One of the most paradoxical features of $\hat{\mathcal{H}}_R$ in the deep-strong-coupling regime is the non-negligible number of photons in the oscillator in the ground state~\cite{Ashhab10PRA}.
In terms of the creation and annihilation operators, the photon number operator is $\hat{a}^{\dagger}\hat{a}$.
Considering the mapping between $\hat{\mathcal{H}}_{circ}$ and $\hat{\mathcal{H}}_{R}$, the photon number operator in $\hat{\mathcal{H}}_{circ}$ is $\hat{\mathcal{H}}_1/(1/\sqrt{L_{LC}C})-0.5$.
As shown in Fig.~\ref{adaga}\textbf{a}, a \erase{non-negligible}\add{nonzero} number of photons in the oscillator is obtained.
Another paradoxical feature of $\hat{\mathcal{H}}_R$ in the deep-strong-coupling regime is a non-negligible expectation value of the field operator $\left \langle \hat{a} + \hat{a}^\dagger\right \rangle$
when the qubit flux bias is $\Phi_{x}\neq 0.5\Phi_0$.
The corresponding operator in $\hat{\mathcal{H}}_{circ}$ is $\hat{\Phi}_1/(L_{LC}I_{zpf})$.
Fig.~\ref{adaga}\textbf{b} shows the expectation values of the fluxes, $\left \langle \hat{\Phi}_1\right \rangle$ and $\left \langle  \hat{\Phi}_2 \right \rangle$, as functions of the flux bias $\Phi_{x}$ for the ground and the first excited states in the case $L_c = 350$~pH.
At $\Phi_{x}/\Phi_0 \neq 0.5$, a nonzero expectation value $\left \langle \hat{\Phi}_1\right \rangle$ is demonstrated.
One may suspect that a nonzero $\left \langle \hat{\Phi}_1\right \rangle$ is unphysical because it might result in a nonzero DC current through the inductor of an LC oscillator in an energy eigenstate.
%Since $a + a^\dagger = \Phi_1/(L_{LC}I_{zpf})$, $\left \langle a + a^\dagger \right \rangle$ is also non-zero at the flux bias $\Phi_{x}\neq 0.5\Phi_0$.
The relation between the flux and current operators is explicitly given in the circuit model [Fig.~\ref{circ}\textbf{a}]:
%Eq: LIPhi
\begin{eqnarray}
\begin{pmatrix}
L_c+L_1 & L_c\\
L_c & L_c+L_2\\
\end{pmatrix}
\begin{pmatrix}
\hat{I}_1\\
\hat{I}_2\\
\end{pmatrix}
=
\begin{pmatrix}
\hat{\Phi}_1\\
\hat{\Phi}_2\\
\end{pmatrix}
,
\label{Eq:LIPhi}
\end{eqnarray}
% Eq: end
where $\hat{I}_k$ ($k = 1,2$) is defined as the operator of the current flowing from the ground node to node $k$.
In the case $L_c = 0$, Eq.~(\ref{Eq:LIPhi}) reduces to $\hat{I}_1=\hat{\Phi}_1/L_1$ and $\hat{I}_2=\hat{\Phi}_2/L_2$ which means that the operator $\hat{\Phi}_1$ can indeed be understood as a current operator.
As shown in Fig.~\ref{adaga}\textbf{c}, at $\Phi_{x}/\Phi_{0} = 0.498$, the expectation value $2\pi \left \langle \hat{\Phi}_1\right \rangle/\Phi_0$ in the ground state is proportional to $L_c$ and is zero only when $L_c=0$.
Fig.~\ref{adaga}\textbf{d} shows the expectation values of the currents $\left \langle \hat{I}_1\right \rangle$ and $\left \langle \hat{I}_2 \right \rangle$ as functions of $\Phi_{x}$ in the case $L_c=350$~pH.
The expectation value $\left \langle \hat{I}_2\right \rangle$ is nonzero at $\Phi_{x}/\Phi_0 \neq 0.5$,
while $\left \langle \hat{I}_1\right \rangle$ is exactly zero at all values of $\Phi_{x}$.
In this way, although $\left \langle \hat{\Phi}_1\right \rangle$ is nonzero in the case $L_c \neq 0$, an unphysical DC current through the inductor of an LC oscillator is not predicted.

%\textcolor{red}{
%\section{Hamiltonian in the charge gauge}
%\label{Hcg}
%}

%\begin{figure}
%\includegraphics{adaga_Lc_cg.pdf}
%\caption{
%(a)~Expectation numbers of photons in the oscillator in the ground state of $\hat{\mathcal{H}}_{circ}'$ as a function of $L_c$ in the charge gauge.
%The solid curve corresponds to $\left \langle 0 \left |\hat{\mathcal{H}}_1'\right | 0 \right \rangle/(\hbar \omega')-0.5$, where $\omega' = 1/\sqrt{L_{LC}}\times\sqrt{(1/C)+(L_{LC}^2/C_JL_{12}^2)}$ is the resonance frequency of $\hat{\mathcal{H}}_1'$.
%(b) [(d)]~Expectation values of fluxes $2\pi \left \langle \hat{\Phi}_k\right \rangle/\Phi_0$ [Expectation values of currents $\left \langle \hat{I}_k \right \rangle$] as functions of the flux bias $\Phi_x$ in the case $L_c=350$~pH.
%Black and red colors respectively indicate $k = 1$ and 2,
%and solid and dashed lines indicate the ground and the excited states, respectively.
%(c)~Expectation value of flux $2\pi \left \langle \hat{\Phi}_1\right \rangle/\Phi_0$ in the ground state of $\hat{\mathcal{H}}_{circ}'$ as a function of $L_c$ at $\Phi_{x}/\Phi_{0} = 0.498$.
%(b)~Expectation values of currents $\left \langle I_1\right \rangle$ and $\left \langle I_2\right \rangle$ as a function of the flux bias $\Phi_x$ for the ground and the first excited states in the case $L_c=350$~pH. 
%}
%\label{exp_cg}
%\end{figure}
Next, we consider the case of $\hat{\mathcal{H}}_{circ}'$.
Since unitary transformations do not change the eigenenergies of Hamiltonians,
$\hat{\mathcal{H}}_{circ}$ and $\hat{\mathcal{H}}_{circ}'$ have exactly the same eigenenergies.
On the other hand, the photon-number and flux operators do not commute with the gauge transformation:
\begin{eqnarray}
\nonumber
\hat{\mathcal{U}}^\dagger \hat{\mathcal{H}}_{1}\hat{\mathcal{U}} & = & \hat{\mathcal{U}}^\dagger \left( \frac{\hat{q}^2_1}{2C} + \frac{\hat{\Phi}^2_1}{2L_{LC}} \right)\hat{\mathcal{U}}\\
&=& \left[ \frac{\hat{q}^2_1}{2C} + \left (\hat{\Phi}_1 +\frac{L_{LC}}{L_{12}}\hat{\Phi}_2\right)^2\frac{1}{2L_{LC}} \right],
\label{UH1U}
\end{eqnarray}
and
\begin{eqnarray}
\hat{\mathcal{U}}^\dagger \hat{\Phi}_1\hat{\mathcal{U}} & = & \hat{\Phi}_1 +\frac{L_{LC}}{L_{12}}\hat{\Phi}_2.
\label{UPhi1U}
\end{eqnarray}

The corresponding photon number operator in $\hat{\mathcal{H}}_{circ}'$ is $\hat{\mathcal{H}}_1'/(\hbar \omega')-0.5$, where $\omega' = 1/\sqrt{L_{LC}}\times\sqrt{(1/C)+(L_{LC}^2/C_JL_{12}^2)}$ is the resonance frequency of $\hat{\mathcal{H}}_1'$.
The corresponding field operator in $\hat{\mathcal{H}}_{circ}'$ is $\hat{\Phi}_1/(L_{LC}I_{zpf})$.
As shown in Fig.~\ref{adaga}\textbf{e}, the expectation number of photons in the charge gauge is much smaller than that in the flux gauge, and the non-negligible expectation number of photons in the oscillator in the ground state arises only in the flux-gauge.
We note here that the resonance frequency of $\hat{\mathcal{H}}_1'$ is larger than $1/\sqrt{L_{LC}C}\simeq 6.03~$GHz:
$\omega'/(2\pi) = 15.66$~GHz in the case $L_c = 350$~pH.
We can see that the expectation value of the operator $\hat{\Phi}_1$ in the case $L_c = 350$~pH is zero at all values of $\Phi_{x}$ [Fig.~\ref{adaga}\textbf{f}], which are also different from the case of the flux-gauge Hamiltonian.
The expectation value of the current operator $\hat{I}_1$ is zero at all values of $\Phi_{x}$, which is true in the flux gauge as well.

%conclusion
\section*{discussion}
\label{discussion}
We have derived the Hamiltonian of a superconducting circuit that comprises a single-Josephson-junction flux qubit and an LC oscillator using the standard quantization procedure.
Excluding the qubit's higher energy levels, the derived circuit Hamiltonian takes the form of the quantum Rabi Hamiltonian.
We show that the Hamiltonian derived from the separate treatment, where the circuit is assumed to be naively divided into the two well-defined components, has the same form as the circuit Hamiltonian, and the inductances in the separate treatment approach those of the circuit Hamiltonian as $L_c$ approaches 0.
%%%
The qubit's higher energy levels mainly cause a negative shift of the entire spectrum, but the energy level structure can still be fitted well by the quantum Rabi Hamiltonian even when the qubit-oscillator circuit is in the deep-strong-coupling regime.
%%%
We also show that although the circuit Hamiltonian can be transformed to a Hamiltonian containing a capacitive coupling term,
the resulting circuit Hamiltonian cannot be approximated by the capacitive-coupling variant of the quantum Rabi Hamiltonian.

\section*{methods}
\label{methods}

As a simple example, let us consider $\hat{\mathcal{H}}_1$ in the main text:
% Eq: Hamiltonian for numerics
\begin{eqnarray}
\nonumber
\hat{\mathcal{H}}_{1} & = & \frac{\hat{q}_1^2}{2C} + \frac{\hat{\Phi}_1^2}{2L_{LC}}\\
&=&4E_C\hat{n}^2 + \frac{1}{2}E_L\hat{\phi}^2,
\label{Eq:H1p}
\end{eqnarray}
% Eq: end
where, $E_C = e^2/(2C)$, $E_L = [\Phi_0/(2\pi)]^2/L_{LC}$,
and $\hat{\phi}$ and $\hat{n}$ are operators of dimensionless magnetic flux and charge, respectively, and satisfy $[\hat{\phi},\hat{n}]=i$.
Using the relation $\hat{\phi} = -(1/i)\partial/\partial n$, the Hamiltonian can be rewritten as
% Eq: Hamiltonian for numerics using partial
\begin{eqnarray}
\hat{\mathcal{H}}_{1} & = & 4E_C\hat{n}^2 - \frac{1}{2}E_L\frac{\partial^2}{\partial n^2}.
\label{Eq:H1pp}
\end{eqnarray}
% Eq: end
Let us calculate wavefunctions $\psi(n)$ and their eigenenergies $E$ of this Hamiltonian.
We expand the wavefunction with plane waves as
% Eq: plane wave expansion
\begin{eqnarray}
\psi(n) = \sum_k \psi_k\frac{\mathrm{e}^{ikn}}{\sqrt{2n_{max}}}.
\label{psin}
\end{eqnarray}
% Eq: end
Here, $2n_{max}$ is the length of the $n$-space.
Considering the periodic boundary condition $\psi(-n_{max}) = \psi(n_{max})$,
the wave number $k$ is given by
% Eq: plane wave expansion
\begin{eqnarray}
k = \frac{2\pi}{2n_{max}}\eta,\,(\eta = 0, \pm1,\pm2,\dots).
\label{k}
\end{eqnarray}
% Eq: end
In numerical calculations in this work, we have used 32 waves for the qubit and 64 waves for the oscillator.
Then, the equation for determining $\psi_k$ and $E$ is obtained as
% Eq: equation transformations
\begin{eqnarray}
\nonumber
\hat{\mathcal{H}}_1\psi(n) & = & E\psi(n)\\
\nonumber
\sum_{k'}\left [4E_Cn^2 + \frac{1}{2}E_Lk'^2 \right ]\psi_{k'}\frac{\mathrm{e}^{ik'n}}{2n_{max}} &=& E\sum_{k'}\psi_{k'}\frac{\mathrm{e}^{ik'n}}{2n_{max}}\\
\nonumber
\int_{-n_{max}}^{n_{max}}dn\frac{\mathrm{e}^{-ikn}}{\sqrt{2n_{max}}}\sum_{k'}\left [ 4E_Cn^2 + \frac{1}{2}E_Lk'^2 \right ]\psi_{k'}\frac{\mathrm{e}^{ik'n}}{2n_{max}} &=& \int_{-n_{max}}^{n_{max}}dn\frac{\mathrm{e}^{-ikn}}{\sqrt{2n_{max}}}E\sum_{k'}\psi_{k'}\frac{\mathrm{e}^{ik'n}}{2n_{max}}\\
\sum_{k'}\left [4E_Cf_{k-k'}(n^2) + \delta_{k,k'}\frac{1}{2}E_Lk^2 \right ]\psi_{k'}&=&E\psi_{k},
\label{HE}
\end{eqnarray}
%Eq: end
where
% Eq: Fourier transform of n^2
\begin{eqnarray}
f_{k-k'}(n^2) \equiv\frac{1}{2n_{max}}\int_{-n_{max}}^{n_{max}}dn \mathrm{e}^{-i(k-k')n}n^2.
\label{Eq:FFT}
\end{eqnarray}
%Eq: end
The set of wavefunctions and eigenenergies are obtained by solving Eq.~(\ref{HE}).
Numerically, we can get $f_{k-k'}(n^2)$ by the fast Fourier transform (FFT) for discretized $n$-space.
After solving the eigenvalue equation in Eq.~(\ref{HE}), the wavefunctions $\psi(n)$ are also obtained from $\psi_k$ by the FFT.
For the calculation of $\hat{\mathcal{H}}_2$ in the main text, we use the following equation instead of Eq.~(\ref{HE}),
\begin{eqnarray}
\sum_{k'}\left \{4E_{CJ}f_{k-k'}(n^2) + \delta_{k,k'}\left [-E_J\cos\left ( k-k_x\right )+\frac{1}{2}E_{LFQ}k^2\right ] \right \}\psi_{k'}&=&E\psi_{k},
\label{HE2}
\end{eqnarray}
where $E_{CJ} = e^2/(2C_J)$, $E_{LFQ} = [\Phi_0/(2\pi)]^2/L_{FQ}$, and $k_x = 2\pi\Phi_x/\Phi_0$.
Note that a fluxonium-resonator circuit that has the identical circuit diagram but different circuit parameters was numerically diagonalized using the harmonic oscillator basis~\cite{Smith2016PRB}.

%%%
%In this way, the circuit Hamiltonian 
%acknowledgment
\begin{acknowledgments}
We are grateful to M. Devoret for valuable discussions.
This work was supported by
Japan Society for the Promotion of Science (JSPS) 
Grants-in-Aid for Scientific Research (KAKENHI) 
(No.~JP19H01831 and JP19K03693),  
Japan Science and Technology Agency (JST)
Precursory Research for Embryonic Science and Technology (PRESTO) (Grant No. JPMJPR1767),
JST Core Research for Evolutionary Science and Technology (CREST)
(Grant No. JPMJCR1775),
and MEXT Quantum Leap Flagship Program (MEXT Q-LEAP) Grant Number JPMXS0120319794.
\end{acknowledgments}

\section*{Author Contributions}
FY conceived the main idea of the paper.
FY, SA, and MB constructed equations of the paper.
FY, SA, TF, and MB conducted numerical calculations.
FY wrote the manuscript with feedback from all authors.
FY and KS supervised the project.

\section*{Supplementary Information}
\setcounter{figure}{0}
\setcounter{equation}{0}
\setcounter{table}{0}
\renewcommand\theequation{S\arabic{equation}}
\renewcommand\thesection{S\arabic{section}}
\renewcommand\thefigure{S\arabic{figure}}
\renewcommand\thetable{S\Roman{table}}

\section{Fitting of $\mathcal{H}_{circ}$ up to the seventh excited state}
\label{seven}

%figure: fitting up to |7>
\begin{figure}
\includegraphics{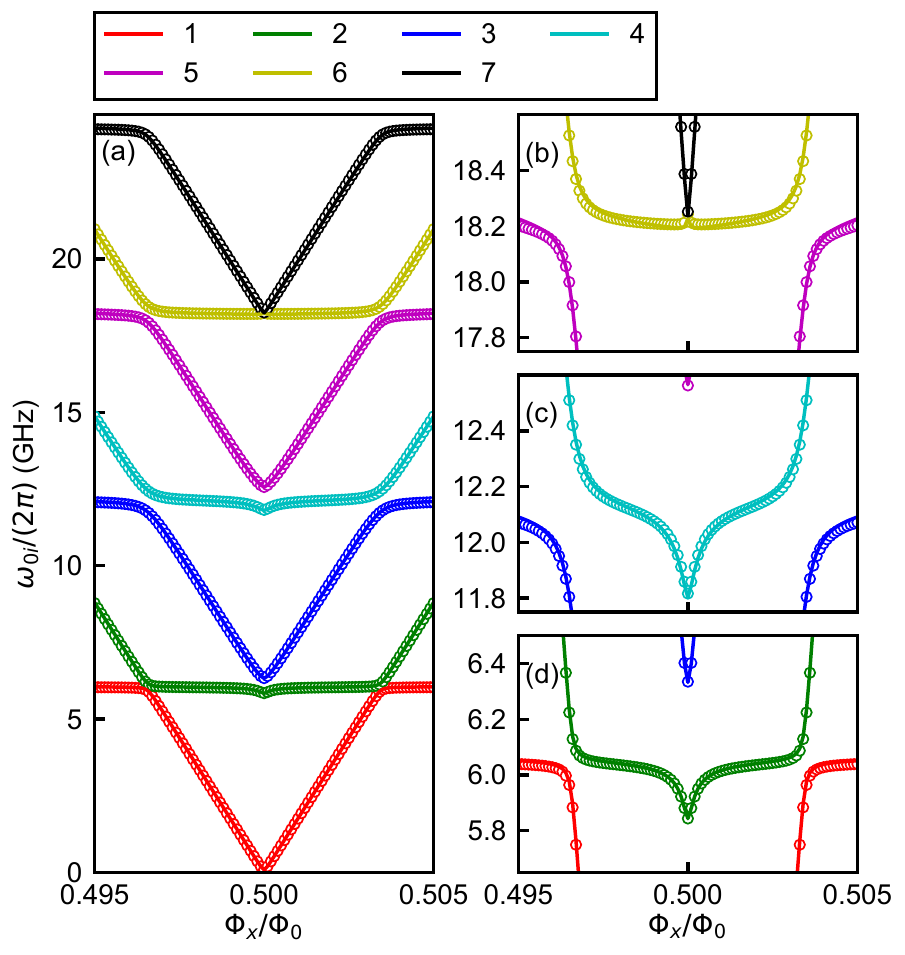}
\caption{
(a)~Transition frequencies of the qubit-oscillator circuit up to the seventh excited state for $L_c = 350$~pH.
Numerically calculated transition frequencies from $\hat{\mathcal{H}}_{circ}$ are plotted as circles, while the results of the fitting by $\hat{\mathcal{H}}_{R}$ are plotted as lines.
The parameters of $\hat{\mathcal{H}}_{R}$ obtained by fitting the spectra of $\hat{\mathcal{H}}_{circ}$ to $\hat{\mathcal{H}}_{R}$ are $\omega/(2\pi) = 6.054$~GHz, $\Delta_q/(2\pi) = 2.133$~GHz, $g/(2\pi)=7.562$~GHz, and $I_p = 282.2$~nA.
Red, green, blue, cyan, magenta, yellow, and black colors indicate transition frequencies $\omega_{01}$, $\omega_{02}$, $\omega_{03}$, $\omega_{04}$, $\omega_{05}$, $\omega_{06}$, and $\omega_{07}$, respectively.
(b), (c), (d)~The same spectra with smaller range of frequency around $3\omega$, $2\omega$, and $\omega$.
}
\label{wij7}
\end{figure}
%end figure
Transition frequencies of the qubit-oscillator circuit numerically calculated from $\hat{\mathcal{H}}_{circ}$ up to the seventh excited state are plotted in Figure~\ref{wij7} for $L_c = 350$~pH\erase{.} \add{using the same circuit parameters as in Fig.~3.}
The calculated spectra can still be fitted well by $\hat{\mathcal{H}}_{R}$.
\add{In the fitting, $\hat{\mathcal{H}}_{R}$ are numerically diagonalized and then the parameters $\omega$, $\Delta_q$, $g$, and $I_p$, are varied to obtain the best fit.}
The average of the squares of the residuals in the least-squares method of obtaining the parameters of $\hat{\mathcal{H}}_{R}$ by fitting,
$\overline{[\delta \omega_{0i}/(2\pi)]^2}$ ($i$ = 1,2,3,4,5,6, and 7), is $152~$MHz$^2$.
This value is almost 10 times larger than the case of the fitting up to the third excited state, but still within a moderate level considering that the largest transition frequency is more than 20~GHz.

\section{Gauge transformation}
\label{GaugeTrans}

The circuit Hamiltonian \add{in the flux gauge} $\hat{\mathcal{H}}_{circ}$ can be written as
\begin{eqnarray}
\nonumber
\hat{\mathcal{H}}_{circ} &=& \frac{\hat{q}_1^2}{2C} + \frac{\hat{q}_2^2}{2C_J}-E_J\cos \left ( 2\pi \frac{\hat{\Phi}_2-\Phi_x}{\Phi_0} \right )+\frac{\hat{\Phi}_1^2}{2L_{LC}} + \frac{\hat{\Phi}_2^2}{2L_{FQ}}-\frac{\hat{\Phi}_1\hat{\Phi}_2}{L_{12}}\\
\nonumber
&=&\frac{\hat{q}_1^2}{2C} + \frac{\hat{q}_2^2}{2C_J}-E_J\cos \left ( 2\pi \frac{\hat{\Phi}_2-\Phi_x}{\Phi_0} \right )\\
&& + \frac{1}{2L_{LC}}\left ( \hat{\Phi}_1-\frac{L_{LC}}{L_{12}}\hat{\Phi}_2\right )^2 + \frac{1}{2}\left ( \frac{1}{L_{FQ}}-\frac{L_{LC}}{L_{12}^2}\right ) \hat{\Phi}_2^2.
\end{eqnarray}
The Hamiltonian can now be transformed into the charge gauge rather easily.
First we note that the unitary operator 
\begin{eqnarray}
\hat{\mathcal{U}} &=& \exp \left ( \frac{1}{i\hbar}\alpha\hat{\Phi}_2\hat{q}_1\right )
\end{eqnarray}
transforms the flux and charge operators as follows:
\begin{eqnarray}
\hat{\mathcal{U}}^\dagger \hat{\Phi}_1 \hat{\mathcal{U}} &=& \hat{\Phi}_1 + \alpha\hat{\Phi}_2,\\
\hat{\mathcal{U}}^\dagger \hat{\Phi}_2 \hat{\mathcal{U}} &=& \hat{\Phi}_2,\\
\hat{\mathcal{U}}^\dagger \hat{q}_1 \hat{\mathcal{U}} &=& \hat{q}_1,\\
\hat{\mathcal{U}}^\dagger \hat{q}_2 \hat{\mathcal{U}} &=& \hat{q}_2 - \alpha\hat{q}_1.
\end{eqnarray}
Then, if we set $\alpha = L_{LC}/L_{12}$, the circuit Hamiltonian is transformed into
\begin{eqnarray}
\nonumber
\hat{\mathcal{H}}_{circ}' &=& \hat{\mathcal{U}}^\dagger \hat{\mathcal{H}}_{circ} \hat{\mathcal{U}}\\
\nonumber
&=&\frac{\hat{q}_1^2}{2C} + \frac{1}{2C_J}\left ( \hat{q}_2 - \frac{L_{LC}}{L_{12}}\hat{q}_1 \right )^2\\
\nonumber
&& -E_J\cos \left ( 2\pi \frac{\hat{\Phi}_2-\Phi_x}{\Phi_0} \right ) + \frac{\hat{\Phi}_1^2}{2L_{LC}} + \frac{1}{2}\left ( \frac{1}{L_{FQ}}-\frac{L_{LC}}{L_{12}^2}\right ) \hat{\Phi}_2^2\\
\nonumber
&=&\left (\frac{1}{2C} + \frac{L_{LC}^2}{2C_JL_{12}^2}\right )\hat{q}_1^2 + \frac{\hat{\Phi}_1^2}{2L_{LC}} + \frac{1}{2C_J}\hat{q}_2^2\\
\label{HcircpAp}
&& + \frac{1}{2}\left ( \frac{1}{L_{FQ}}-\frac{L_{LC}}{L_{12}^2}\right ) \hat{\Phi}_2^2
 - E_J\cos \left ( 2\pi \frac{\hat{\Phi}_2-\Phi_x}{\Phi_0} \right ) - \frac{L_{LC}}{C_JL_{12}}\hat{q}_1\hat{q}_2.
\end{eqnarray}

\section{Energy shifts up to second order in perturbation theory}
\label{perturb}

To investigate the reason for the poor reproducibility of the characteristic spectra of $\hat{\mathcal{H}}_{circ}'$ by $\hat{\mathcal{H}}_R'$, we calculate the energy shifts of the four lowest energy levels using perturbation theory.
Let us consider the Hamiltonian $\hat{\mathcal{H}}_{circ,\lambda} \equiv \hat{\mathcal{H}}_{1} + \hat{\mathcal{H}}_{2} + \lambda\hat{\mathcal{H}}_{12}$.
Here, we assume that the eigenenergies and the eigenstates of $\hat{\mathcal{H}}_{circ,\lambda}$ can be described by \add{a} power series \add{in}\erase{of} $\lambda$ as $E_{ni,\lambda} = \sum_{k=0}^{\infty}\lambda^kE_{ni}^{(k)}$ and $\left | ni \right \rangle_\lambda = \sum_{k=0}^{\infty}\lambda^k\left | ni^{(k)} \right \rangle$,
where $E_{ni}^{(0)}$ and $\left | ni^{(0)}\right \rangle$ are the eigenenergies and the eigenstates of the non-interacting Hamiltonian $\hat{\mathcal{H}}_{1} + \hat{\mathcal{H}}_{2}$,
$n$ is the number of photons in the oscillator, and $i=g,e$ represents the eigenstate of $\hat{\mathcal{H}}_{2}$.
By taking the eigenenergies and the eigenstates of $\hat{\mathcal{H}}_{circ,\lambda}$ and multiplying by $\left \langle mj^{(k)}\right |$ from the left,
the correction terms can be written as
\begin{eqnarray}
E_{ni}^{(1)} &=& \left \langle ni^{(0)} \right | \hat{\mathcal{H}}_{12} \left | ni^{(0)}\right \rangle
\end{eqnarray}
and
\begin{eqnarray}
E_{ni}^{(2)} &=& \sum_{m,j} \frac{\left | \left \langle mj^{(0)} \left | \hat{\mathcal{H}}_{12} \right | ni^{(0)}\right \rangle \right |^2}{E_{ni}^{(0)}-E_{mj}^{(0)}}\equiv\sum_{m,j} \hbar\chi_{ni,mj}.
\label{Eni2}
\end{eqnarray}
Here, $\chi_{ni,mj}$ is the energy shift of the state $\left | ni\right \rangle$ due to the interaction with the state $\left | mj\right \rangle$, where $m = 0,1,2,\dots$, $j = g,e,f,h,\dots$, and $\left | f\right \rangle$ and $\left | h\right \rangle$ respectively represent the second and \erase{the} third excited states of $\hat{\mathcal{H}}_{2}$. 
We find that $E_{ni}^{(1)}=0$ for all the combinations of $n = 0, 1$ and $i = g, e$,
and the total energy shifts up to the second-order perturbation $\chi_{ni}$ are simply determined by the second-order correction terms: $\chi_{ni} = \sum_{m,j} \chi_{ni,mj}$.
We also find that the third- or higher-order correction terms are much smaller than the second-order correction term:
the numerator of the $n$th order correction term is the product of $n$ matrix elements of $\hat{H}_{12}$, which are at most several hundred MHz, while the denominator of the $n$th order correction term is the product of $n-1$ energy-level differences, which are in the order of GHz or more.

\begin{figure*}
\includegraphics{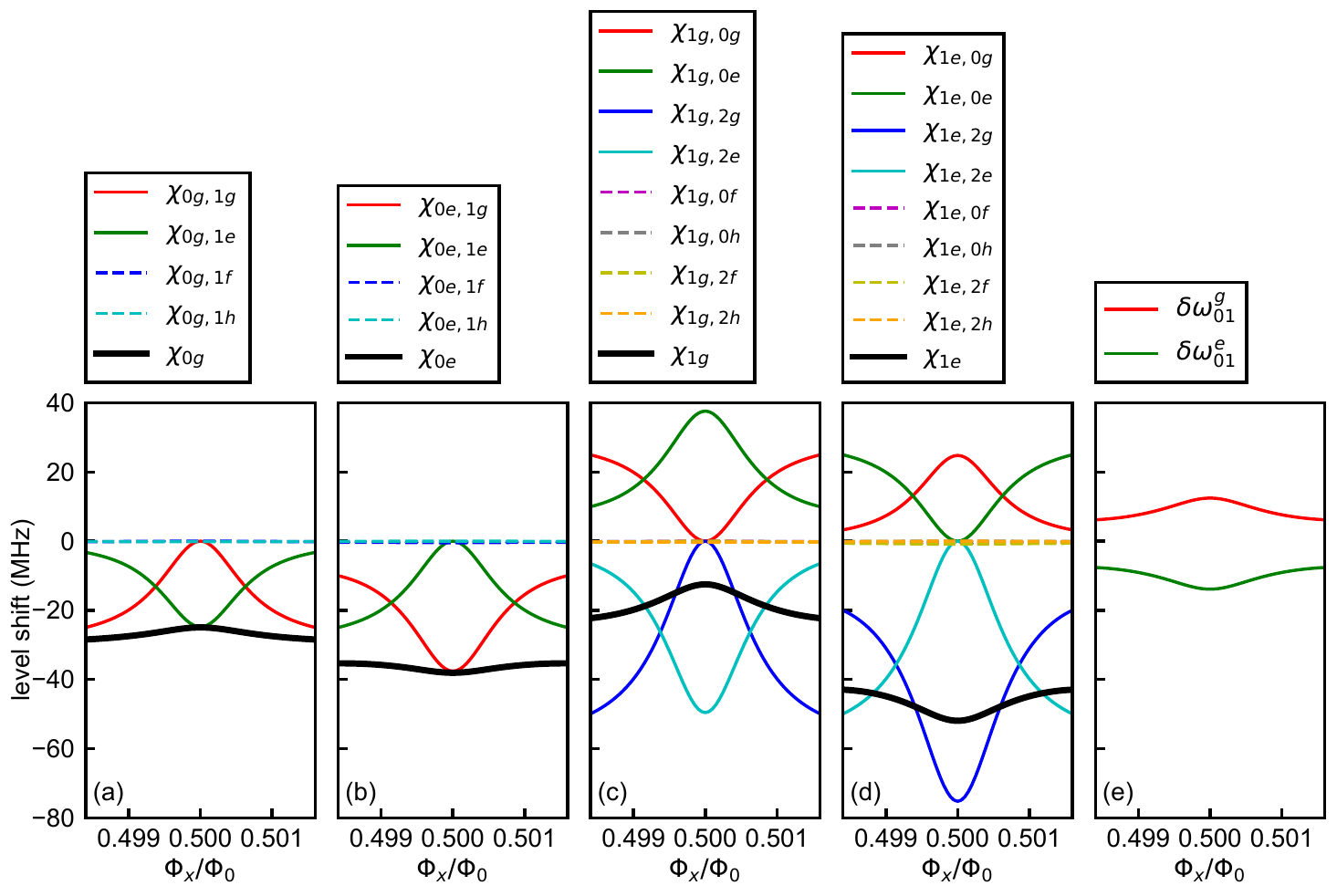}
\caption{
Nonzero energy level shifts $\chi_{ni,mj}$ of the non-interacting Hamiltonian $\hat{\mathcal{H}}_{1} + \hat{\mathcal{H}}_{2}$ due to the term $\hat{\mathcal{H}}_{12}$ and the state $\left | mj\right \rangle$ for the four lowest eigenstates $\left | ni\right \rangle=$ (a)~$\left | 0g\right \rangle$, (b)~$\left | 0e\right \rangle$, (c)~$\left | 1g\right \rangle$, and (d)~$\left | 1e\right \rangle$ for $L_c = 20$~pH.
(e)~the net change of the transition frequencies $\delta \omega_{01}^g = \chi_{1g} - \chi_{0g}$ and $\delta \omega_{01}^e = \chi_{1e} - \chi_{0e}$, obtained by combining the shifts in Panels~(a)-(d).
}
\label{chiF}
\end{figure*}

Figures~\ref{chiF}(a)-(d) show nonzero energy level shifts $\chi_{ni,mj}$ of the non-interacting Hamiltonian $\hat{\mathcal{H}}_{1} + \hat{\mathcal{H}}_{2}$ due to the interaction term $\hat{\mathcal{H}}_{12}$ for the lowest four eigenstates $\left | ni\right \rangle = \left | 0g\right \rangle$, $\left | 0e\right \rangle$, $\left | 1g\right \rangle$, and $\left | 1e\right \rangle$ of the non-interacting Hamiltonian $\hat{\mathcal{H}}_{1} + \hat{\mathcal{H}}_{2}$ for $L_c = 20$~pH.
Although some energy shifts $\chi_{ni,mj}$ appear to be equal to zero everywhere, they are in fact finite but extremely small.
The level shifts $\chi_{ni,mj}$ that involve the higher qubit levels $j = f, h$ plotted using dashed lines are all close to zero, and the net level shifts are almost completely determined by the eigenstates involving the two lowest energy levels of $\hat{\mathcal{H}}_{2}$.
Figure~\ref{chiF}(e) shows the net change of the transition frequencies $\delta \omega_{01}^g = \chi_{1g} - \chi_{0g}$ and $\delta \omega_{01}^e = \chi_{1e} - \chi_{0e}$ from the resonance frequency of the oscillator described by $\hat{\mathcal{H}}_{1}$,
%\add{Although the quantitative agreement is still not excellent,}
which reproduce the
%, which is the corresponding transition frequencies of the non-interacting Hamiltonian $\hat{\mathcal{H}}_{1} + \hat{\mathcal{H}}_{2}$.
peaks and dips \add{that occur} in the spectrum at $\Phi_x/\Phi_0 = 0.5$ shown in Fig.~4\textbf{a} in the main text.
%This result is consistent with the fact that $\hat{\mathcal{H}}_{R}$ accurately reproduces the transition frequencies calculated from $\hat{\mathcal{H}}_{circ}$ as shown in Fig.~\ref{wij_phix}(a).
\begin{figure*}
\includegraphics{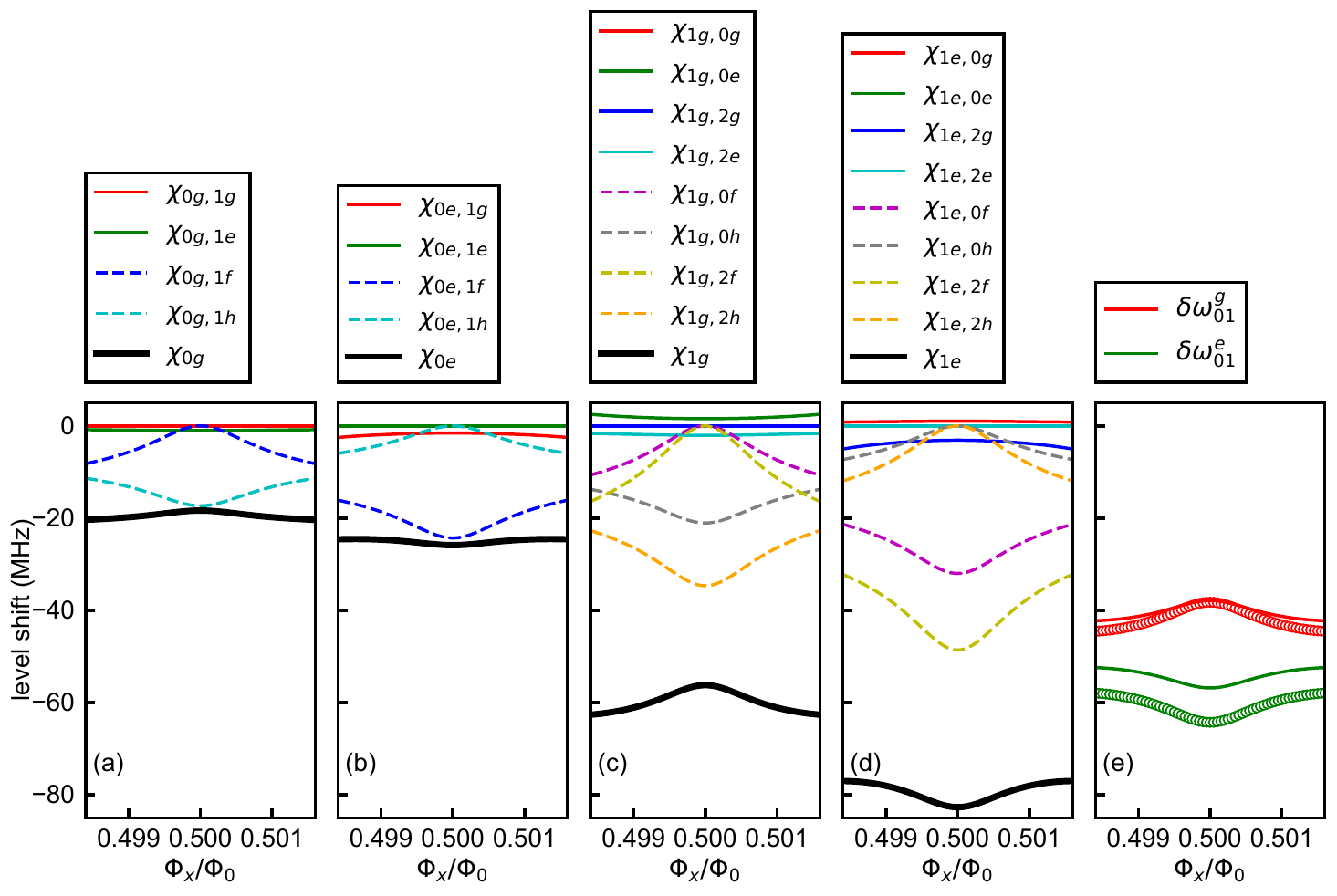}
\caption{Nonzero energy level shifts $\chi_{ni,mj}$ of the non-interacting Hamiltonian $\hat{\mathcal{H}}_{1}' + \hat{\mathcal{H}}_{2}'$ due to the term $\hat{\mathcal{H}}_{12}'$ and the state $\left | mj\right \rangle$ for the four lowest eigenstates $\left | ni\right \rangle=$ (a)~$\left | 0g\right \rangle$, (b)~$\left | 0e\right \rangle$, (c)~$\left | 1g\right \rangle$, and (d)~$\left | 1e\right \rangle$ for $L_c = 20$~pH.
(e)~the net changes of the transition frequencies $\delta \omega_{01}^g = \chi_{1g} - \chi_{0g}$ and $\delta \omega_{01}^e = \chi_{1e} - \chi_{0e}$, obtained by combining the shifts in Panels~(a)-(d).
\add{Numerically calculated transition frequencies of $\hat{\mathcal{H}}_{circ}$, $\omega_{02}-\omega'$ and $\omega_{13}-\omega'$, are plotted as circles, where $\omega'$ is the resonance frequency of $\hat{\mathcal{H}}'_{circ}$.}
}
\label{chiC}
\end{figure*}

We now consider the case of the circuit Hamiltonian $\hat{\mathcal{H}}_{circ}'$.
Figures~\ref{chiC}(a)-(d) show nonzero energy level shifts $\chi_{ni,mj}$ of the non-interacting Hamiltonian $\hat{\mathcal{H}}_{1}' + \hat{\mathcal{H}}_{2}'$ due to the interaction with the state $\left | mj\right \rangle$ for eigenstates $\left | ni\right \rangle$ ($n = 0, 1$, $i = g, e$) for $L_c = 20$~pH.
Figure~\ref{chiC}(e) shows the net change of the transition frequencies $\delta \omega_{01}^g = \chi_{1g} - \chi_{0g}$ and $\delta \omega_{01}^e = \chi_{1e} - \chi_{0e}$ from the resonance frequency of the oscillator described by $\hat{\mathcal{H}}'_{1}$.
\add{Although the qualitative agreement is still not excellent, the peaks and dips that occur in the spectrum at $\Phi_x/\Phi_0$ are qualitatively reproduced.
The agreement of $\delta \omega_{01}^g$ is better than that of $\delta \omega_{01}^e$.}
%Although some of energy shifts $\chi_{\left |ni\right \rangle \left | mj\right \rangle}$ look staying at zero, they are just close to zero but nonzero.
%Two points are worth mentioning:
Contrary to the case of the circuit Hamiltonian $\hat{\mathcal{H}}_{circ}$, the level shifts $\chi_{ni,mj}$ that involve the higher qubit levels $j = f, h$ plotted using dashed lines are larger than those including $j = g, e$ in most of the flux bias conditions, and the level shifts are mainly determined by the third and the fourth lowest energy levels of $\hat{\mathcal{H}}_{2}'$.
%These results explain the reason for the poor reproducibility of the characteristic spectra of $\hat{\mathcal{H}}_{circ}'$ by $\hat{\mathcal{H}}_R'$, in which $\hat{\mathcal{H}}'_{FQ}$ contains only the lowest two energy levels as shown in Fig.~\ref{wij_cg}(a):
These results explain the difference between the spectra of $\hat{\mathcal{H}}_{circ}'$ and those of $\hat{\mathcal{H}}_R'$, in which $\hat{\mathcal{H}}'_{FQ}$ contains only the lowest two energy levels, as shown in Fig.~4\textbf{e} in the main text:
%We note here that the changes of the transition frequencies $\delta \omega_{01}^g$ and $\delta \omega_{01}^e$ from the resonance frequency of the oscillator described by $\hat{\mathcal{H}}'_{1}$ are both negative, which compensate the increase of the resonance frequency as can be seen in Figs.~\ref{wij_cg}(a) and \ref{wij_cg}(b).
The spectrum represented by solid grey and red lines obtained from $\hat{\mathcal{H}}_{R}'$ are similar to those of an uncoupled qubit-oscillator circuit, whereas the spectrum represented by the open grey and red circles obtained from $\hat{\mathcal{H}}_{circ}'$ show the peaks and dips around the symmetry point and the overall frequency is lower.
%These differences are consistent with the changes of the transition frequencies $\delta \omega_{01}^g$ and $\delta \omega_{01}^e$.
%If we take the negative shift of the resonance frequency into account, the expectation numbers of photons defined as $\left \langle 0 \left |\hat{\mathcal{H}}_1'\right | 0 \right \rangle/(\hbar \omega_1')-0.5$ becomes larger: 15.66/6.03 $\sim$2.6 times larger for $L_c=350$~pH.
%Nevertheless, the expectation number of photons in the charge gauge is much smaller than that in the flux gauge.

\section{Matrix elements of the charge and flux operators}
\label{Mij}

In Section~\ref{perturb}, the energy level shifts of the non-interacting Hamiltonian caused by the interaction term were studied using perturbation theory.
The energy shifts of the states $\left | ni \right \rangle$ are given in Eq.~(\ref{Eni2}).
Considering that the states $\left | ni^{(0)}\right \rangle$ and $\left | mj^{(0)}\right \rangle$ are separable, the following matrix elements can be written as products of matrix elements of the oscillator and the qubit:
\begin{eqnarray}
\left \langle mj^{(0)}\left | \hat{\Phi}_1\hat{\Phi}_2 \right | ni^{(0)}\right \rangle = \left \langle m\left | \hat{\Phi}_1 \right | n\right \rangle\left \langle j\left | \hat{\Phi}_2 \right | i\right \rangle
\end{eqnarray}
and
\begin{eqnarray}
\left \langle mj^{(0)}\right | \hat{q}_1\hat{q}_2 \left | ni^{(0)}\right \rangle = \left \langle m\left | \hat{q}_1 \right | n\right \rangle\left \langle j\left | \hat{q}_2 \right | i\right \rangle.
\end{eqnarray}
The matrix elements of the oscillator operators $\hat{\Phi}_1$ and $\hat{q}_1$ are analytically given as
\begin{eqnarray}
\nonumber
\left \langle m\left | \hat{\Phi}_1 \right | n\right \rangle &=& L_{LC}I_{zpf}\left \langle m\left | (\hat{a} + \hat{a}^\dagger ) \right | n\right \rangle\\
&=&L_{LC}I_{zpf}(\sqrt{n}\delta_{m,n-1} + \sqrt{n+1}\delta_{m,n+1})
\end{eqnarray}
and
\begin{eqnarray}
\nonumber
\left \langle m\left | \hat{q}_1 \right | n\right \rangle &=& -iCV_{zpf}\left \langle m\left | (\hat{a} - \hat{a}^\dagger ) \right | n\right \rangle\\
&=&-iCV_{zpf}(\sqrt{n}\delta_{m,n-1} - \sqrt{n+1}\delta_{m,n+1}).
\end{eqnarray}
Only matrix elements that satisfy $m = n\pm 1$ are nonzero.
The matrix elements of the qubit operators $\hat{\Phi}_2$ and $\hat{q}_2$ for $i = g, e$ and $j = g, e, f, h, k, l$, where $\left | k \right \rangle$ and $\left | l \right \rangle$ respectively represent the fourth and\erase{ the} fifth excited states of $\hat{\mathcal{H}}_2$, are numerically calculated as shown in Fig.~6 in the main text.
Note that the matrix elements of the qubit operators at $\varepsilon = 0$ and with a quadratic-plus-quartic potential energy function were studied in Ref. \cite{Bernardis2018PRA}.

Regarding the matrix elements $\left | \left \langle j\left | \hat{\Phi}_2 \right | i\right \rangle \right |$ ($i = g,e$), those involving the higher qubit levels $j = f, h, k, l$ are smaller than those of $j = g, e$.
Together with the fact that the energy difference $\left | E_{ni}^{(0)} - E_{mj}^{(0)}\right |$ of $j = f, h, k, l$ is larger than those of $j = g, e$, which appears in the denominator of Eq.~(\ref{Eni2}), this result explains why the level shifts are almost completely determined by the eigenstates involving the two lowest energy levels of $\hat{\mathcal{H}}_2$.
Regarding the matrix elements $\left | \left \langle j\left | \hat{q}_2 \right | i\right \rangle \right |$ ($i = g,e$) on the other hand, some of those involving the higher qubit levels $j = f, h, k, l$ are significantly larger than those of $j = g, e$ in most of the flux bias range.
Although the energy difference $\left | E_{ni}^{(0)} - E_{mj}^{(0)}\right |$ is larger than those of $j = g, e$, the large matrix elements result in a situation where the level shifts are mainly determined by the third and\erase{ the} fourth lowest energy levels of $\hat{\mathcal{H}}'_2$.
We note here that the matrix elements involving the levels $j = k,l$ are smaller than those involving the levels $j = f,h$ and the energy difference is larger than those of $j = f,h$.
Hence, the level shifts caused by higher levels with $j = k,l$ are smaller than those with $j = f,h$.

The relation between the different matrix elements can be intuitively understood using an analogy to a basic quantum physics problem.
The qubit Hamiltonians $\hat{\mathcal{H}}_2$ and $\hat{\mathcal{H}}'_2$ have the same form as the Hamiltonian of a single particle in a trapping potential, with $\hat{\Phi}_2$ and $\hat{q}_2$ playing the roles of the position and momentum variables, respectively. 
Around the symmetry point, the trapping potential is a double-well potential.
The energy eigenstates in the double-well potential are superpositions of the energy eigenstates in the two separate wells, which are approximately harmonic oscillator potentials.
We therefore have superpositions of harmonic oscillator states whose centers are separated by a distance that is larger than their widths.
\begin{figure*}
\includegraphics{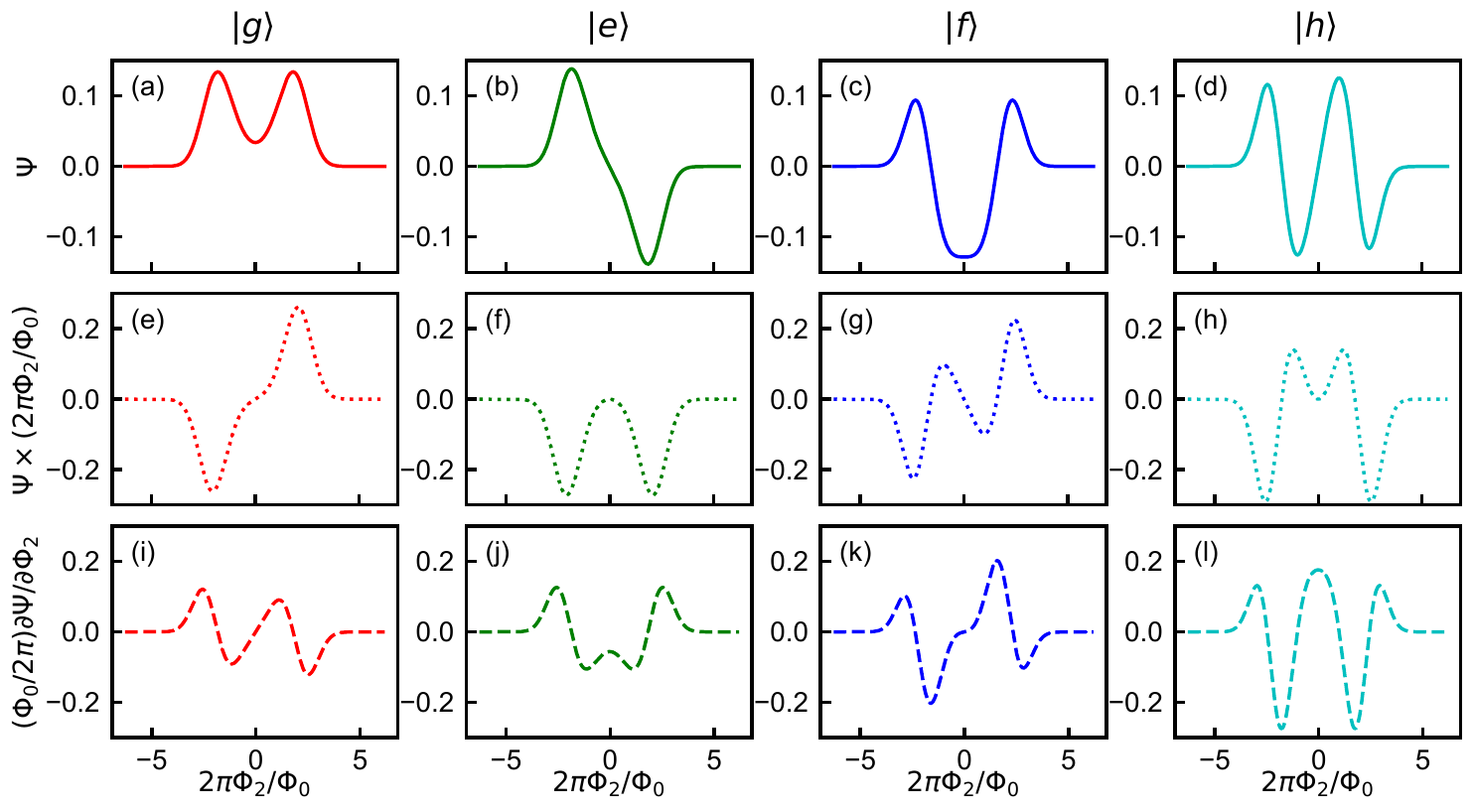}
\caption{Numerically calculated wavefunctions of the eigenstates of $\hat{\mathcal{H}}_2$, $\Psi_i$, as well as the combinations $\Psi_i \times (2\pi \Phi_2/\Phi_0)$, and $(\Phi_0/2\pi \Phi_2)\partial \Psi_i/\partial \Phi_2$ $(i = g, e, f, h)$, as functions of $2\pi \Phi_2/\Phi_0$ for $\Phi_x/\Phi_0 = 0.5$.
}
\label{wfs}
\end{figure*}
Figures~\ref{wfs}(a)-(d) show wave functions of the eigenstates of $\hat{\mathcal{H}}_2$, $\Psi_{i}$ $(i = g, e, f, h)$, as functions of $\Phi_2$ at the flux bias point $\Phi_x/\Phi_0 = 0.5$.
Matrix elements of the position operator can be written using the wave-function representation:
\begin{eqnarray}
\left \langle j\left | \hat{\Phi}_2 \right | i\right \rangle = \int d\Phi_2 (\Psi_j^* \Phi_2 \Psi_i).
\end{eqnarray}
Matrix elements that combine a pair of functions $\Psi_j$ and $\Phi_2\Psi_i$ [Figs.~\ref{wfs}(e)-(h)] with similar shapes are large.
At the symmetry point, the matrix elements $\left | \left \langle e\left | \hat{\Phi}_2 \right | g\right \rangle \right |$ = $\left | \left \langle g\left | \hat{\Phi}_2 \right | e\right \rangle \right |$ $\sim$ $\left | \left \langle h\left | \hat{\Phi}_2 \right | f\right \rangle \right |$ = $\left | \left \langle f\left | \hat{\Phi}_2 \right | h\right \rangle \right |$ are large and the others are small or zero.
Away from the symmetry point, the matrix elements $\left | \left \langle g\left | \hat{\Phi}_2 \right | g\right \rangle \right |$, $\left | \left \langle e\left | \hat{\Phi}_2 \right | e\right \rangle \right |$, $\left | \left \langle f\left | \hat{\Phi}_2 \right | f\right \rangle \right |$, and $\left | \left \langle h\left | \hat{\Phi}_2 \right | h\right \rangle \right |$ also become large.
The matrix elements between states with $i = g,e$ and states with $j = f,h$ are therefore small compared to matrix elements involving only $g$ and $e$.
%%%
Similarly, the matrix elements of the momentum operator can be written using the wave-function representation:
\begin{eqnarray}
\left \langle j\left | \hat{q}_2 \right | i\right \rangle = \int d\Phi_2 \left (\Psi_j^*\frac{\hbar}{i}\frac{\partial \Psi_i}{\partial \Phi_2}\right ).
\end{eqnarray}
Matrix elements that have a pair of functions $\Psi_j$ and $\partial \Psi_i/\partial \Phi_2$ [Figs.~\ref{wfs}(i)-(l)] with similar shapes are large.
At the symmetry point, the matrix elements $\left | \left \langle h\left | \erase{\hat{\Phi}_2} \add{\hat{q}_2} \right | g\right \rangle \right |$ = $\left | \left \langle g\left | \erase{\hat{\Phi}_2} \add{\hat{q}_2} \right | h\right \rangle \right |$ $\sim$ $\left | \left \langle f\left | \erase{\hat{\Phi}_2} \add{\hat{q}_2} \right | e\right \rangle \right |$ = $\left | \left \langle e\left | \erase{\hat{\Phi}_2} \add{\hat{q}_2} \right | f\right \rangle \right |$ are large and the others are small or zero.
Away from the symmetry point, the matrix elements $\left | \left \langle f\left | \erase{\hat{\Phi}_2} \add{\hat{q}_2} \right | g\right \rangle \right |$, $\left | \left \langle g\left | \erase{\hat{\Phi}_2} \add{\hat{q}_2} \right | f\right \rangle \right |$, $\left | \left \langle h\left | \erase{\hat{\Phi}_2} \add{\hat{q}_2} \right | e\right \rangle \right |$, and $\left | \left \langle e\left | \erase{\hat{\Phi}_2} \add{\hat{q}_2} \right | h\right \rangle \right |$ also become large.
The matrix elements between states with $i = g,e$ and states with $j = f, h$ are therefore large compared to matrix elements involving only $g$ and $e$.

%%% use BibTeX %%%
%\bibliographystyle{apsrev4-1}
%\bibliography{fy}
%\end{document}
%%% use BibTeX %%%

%merlin.mbs apsrev4-1.bst 2010-07-25 4.21a (PWD, AO, DPC) hacked
%Control: key (0)
%Control: author (72) initials jnrlst
%Control: editor formatted (1) identically to author
%Control: production of article title (-1) disabled
%Control: page (0) single
%Control: year (1) truncated
%Control: production of eprint (0) enabled
%

%
\end{document}